\documentclass[aps,prc,superscriptaddress,showpacs,nofootinbib,floatfix,onecolumn]{revtex4}
\usepackage{atlasphysics} % Contains useful shortcuts. Uncomment to use
\usepackage[colorlinks,linkcolor=blue,citecolor=blue,filecolor=black,urlcolor=blue,pdfauthor={The ATLAS Collaboration}]{hyperref}
\usepackage{preprintcover}  % load package
\PreprintCoverPaperTitle{\boldmath Measurement of the correlation between flow harmonics of different order in lead--lead collisions at $\sqrt{s_{_{N\!N}}}$=2.76 TeV with the ATLAS detector}
\PreprintIdNumber{CERN-PH-EP-2015-079}  % CERN preprint number

\PreprintCoverAbstract{Correlations between the elliptic or triangular flow coefficients $v_m$ ($m=2$ or 3) and other flow harmonics $v_n$ ($n=2$ to 5) are measured using $\sqrt{s_{_{N\!N}}}=2.76$ TeV Pb+Pb collision data collected in 2010 by the ATLAS experiment at the LHC, corresponding to an integrated lumonisity of 7 $\mu$b$^{-1}$. The $v_m$--$v_n$ correlations are measured in midrapidity as a function of centrality, and, for events within the same centrality interval, as a function of event ellipticity or triangularity defined in a forward rapidity region. For events within the same centrality interval, $v_3$ is found to be anticorrelated with $v_2$ and this anticorrelation is consistent with similar anticorrelations between the corresponding eccentricities $\epsilon_2$ and $\epsilon_3$. On the other hand, it is observed that $v_4$ increases strongly with $v_2$, and $v_5$ increases strongly with both $v_2$ and $v_3$. The trend and strength of the $v_m$--$v_n$ correlations for $n=4$ and 5 are found to disagree with $\epsilon_m$--$\epsilon_n$ correlations predicted by initial-geometry models. Instead, these correlations are found to be consistent with the combined effects of a linear contribution to $v_n$ and a nonlinear term that is a function of $v_2^2$ or of $v_2v_3$, as predicted by hydrodynamic models. A simple two-component fit is used to separate these two contributions. The extracted linear and nonlinear contributions to $v_4$ and $v_5$ are found to be consistent with previously measured event-plane correlations.}
\PreprintJournalName{Phys. Rev. C.}  % journal name

\usepackage{afterpage}
\usepackage{setspace}
\usepackage{rotating}
\usepackage{amssymb}
\usepackage{float}
\usepackage{harpoon}
\usepackage{amsmath}

\begin{document}
\newcommand{\Dphi}{\mbox{$\Delta \phi$}}
\newcommand{\Deta}{\mbox{$\Delta \eta$}}
\newcommand{\ptref}{\mbox{$p_{\mathrm{T}}^{\mathrm{ref}}$}}
\newcommand{\pta}{\mbox{$p_{\mathrm{T}}^{\mathrm{a}}$}}
\newcommand{\ptb}{\mbox{$p_{\mathrm{T}}^{\mathrm{b}}$}}
\newcommand{\ptab}{\mbox{$p_{\mathrm{T}}^{\mathrm{a,b}}$}}
\newcommand{\npart}{\mbox{$N_{\mathrm{part}}$}}
\title{Measurement of the correlation between flow harmonics of different order in lead--lead collisions at $\sqrt{s_{_{N\!N}}}$=2.76 TeV with the ATLAS detector}
\author{The ATLAS Collaboration}
\begin{abstract}
Correlations between the elliptic or triangular flow coefficients $v_m$ ($m=2$ or 3) and other flow harmonics $v_n$ ($n=2$ to 5) are measured using $\sqrt{s_{_{N\!N}}}=2.76$ TeV Pb+Pb collision data collected in 2010 by the ATLAS experiment at the LHC, corresponding to an integrated lumonisity of 7 $\mu$b$^{-1}$. The $v_m$--$v_n$ correlations are measured in midrapidity as a function of centrality, and, for events within the same centrality interval, as a function of event ellipticity or triangularity defined in a forward rapidity region. For events within the same centrality interval, $v_3$ is found to be anticorrelated with $v_2$ and this anticorrelation is consistent with similar anticorrelations between the corresponding eccentricities $\epsilon_2$ and $\epsilon_3$. On the other hand, it is observed that $v_4$ increases strongly with $v_2$, and $v_5$ increases strongly with both $v_2$ and $v_3$. The trend and strength of the $v_m$--$v_n$ correlations for $n=4$ and 5 are found to disagree with $\epsilon_m$--$\epsilon_n$ correlations predicted by initial-geometry models. Instead, these correlations are found to be consistent with the combined effects of a linear contribution to $v_n$ and a nonlinear term that is a function of $v_2^2$ or of $v_2v_3$, as predicted by hydrodynamic models. A simple two-component fit is used to separate these two contributions. The extracted linear and nonlinear contributions to $v_4$ and $v_5$ are found to be consistent with previously measured event-plane correlations.
\end{abstract}
\maketitle

\section{Introduction}

Heavy-ion collisions at the Relativistic Heavy Ion Collider (RHIC) and the Large Hadron Collider (LHC) create hot and dense matter that is thought to be composed of strongly coupled quarks and gluons. The distribution of this matter in the transverse plane is both non-uniform in density and asymmetric in shape~\cite{Gyulassy:1996br,Alver:2006wh}. The matter expands under large pressure gradients, which transfer the inhomogenous initial condition into azimuthal anisotropy of produced particles in momentum space~\cite{Ollitrault:1992bk,Alver:2010gr}. Hydrodynamic models are used to understand the space-time evolution of the matter by comparing predictions with the measured azimuthal anisotropy~\cite{Gale:2013da,Heinz:2013th,Luzum:2013yya}. The success of these models in describing the anisotropy of particle production in heavy-ion collisions at RHIC and the LHC~\cite{Adare:2011tg,Adamczyk:2013waa,ALICE:2011ab,Aad:2012bu,Chatrchyan:2013kba,Aad:2013xma,Aad:2014fla} places significant constraints on the transport properties (such as the ratio of shear viscosity to entropy density) and initial conditions of the produced matter~\cite{Luzum:2012wu,Teaney:2010vd,Gale:2012rq,Niemi:2012aj,Qiu:2012uy,Teaney:2013dta}.

The azimuthal anisotropy of the particle production in each event can be characterized by a Fourier expansion of the corresponding probability distribution $\emph{P}(\phi)$ in azimuthal angle $\phi$~\cite{Ollitrault:1992bk,Voloshin:1994mz}, 
\begin{eqnarray}
\label{eq:1}
\emph{P}(\phi) = \frac{1}{2\pi}\left[1+ \sum_{n=1}^{\infty} \left({\bf v}_{n}e^{-in\phi}+\left[{\bf v}_{n}e^{-in\phi}\right]^*\right)\right],\;\;\; {\bf v}_n = v_n e^{in\Phi_n},
\end{eqnarray}
where $v_n$ and $\Phi_n$ are the magnitude and phase (also known as the event plane or EP) of the $n^{\mathrm{th}}$-order harmonic flow, and $\emph{P}(\phi)$ is real by construction. The presence of harmonic flow has been related to various moments of shape configurations of the initially produced fireball. These moments are described by the eccentricity vector ${\boldsymbol \epsilon}_n$ calculated from the transverse positions $(r,\phi)$ of the participating nucleons relative to their center of mass~\cite{Alver:2010gr,Teaney:2010vd}:
\begin{eqnarray}
\label{eq:0c}
{\boldsymbol \epsilon}_n= \epsilon_n e^{i n\Psi_n} = -\frac{\langle r^n e^{i n\phi} \rangle}{\langle r^{n}\rangle}\;,
\end{eqnarray} 
where $\langle...\rangle$ denotes an average over the transverse position of all participating nucleons, and $\epsilon_n$ and $\Psi_n$ (also known as the participant plane or PP) represent the magnitude and orientation of the eccentricity vector, respectively. The eccentricity vectors characterize the spatial anisotropy of the initially produced fireball, which drives the flow harmonics in the final state.

According to hydrodynamic model calculations, elliptic flow ${\bf v}_2$ and triangular flow ${\bf v}_3$ are the dominant harmonics, and they are driven mainly by the ellipticity vector ${\boldsymbol \epsilon}_2$ and triangularity vector ${\boldsymbol \epsilon}_3$ of the initially produced fireball~\cite{Qiu:2011iv,Gardim:2011xv}:
\begin{eqnarray}
\label{eq:0d}
v_2e^{i2\Phi_2} \propto \epsilon_2e^{i2\Psi_2},\;\; v_3e^{i3\Phi_3} \propto \epsilon_3e^{i3\Psi_3}\;.
\end{eqnarray}
This proportionality is often quantified by a ratio
\begin{eqnarray}
\label{eq:0dd}
k_n = v_n/\epsilon_n,\;\; n=2\;\; \mathrm{or}\;\;3,
\end{eqnarray}
where the linear response coefficients $k_n$ are found to be independent of the magnitude of $\epsilon_n$ but change with centrality~\cite{Voloshin:2008dg,Qiu:2011iv}.

The origin of higher-order ($n>3$) harmonics is more complicated; they arise from both ${\boldsymbol\epsilon}_n$ and nonlinear mixing of lower-order harmonics~\cite{Gardim:2011xv,Teaney:2012ke,Teaney:2013dta}. For example, an analytical calculation shows that the $v_4$ signal comprises a term proportional to $\epsilon_4$ (linear response term) and a leading nonlinear term that is proportional to $\epsilon_2^2$~\cite{Gardim:2011xv,Teaney:2012gu}:
\begin{eqnarray}
\nonumber
v_4e^{i4\Phi_4} &=& a_{0}\; \epsilon_4e^{i4\Psi_4}  + a_{1}\; \left(\epsilon_2e^{i2\Psi_2}\right)^{2} +...\;\\\label{eq:0e}
              &= & c_0\;  e^{i4\Psi_4}  + c_{1}\; \left(v_2e^{i2\Phi_2}\right)^{2} +...\;,
\end{eqnarray}
where the second line of the equation follows from Eq.~(\ref{eq:0d}), and $c_0=a_{0}\epsilon_4$ denotes the linear component of $v_4$ and coefficients $a_0$, $a_1$ and $c_1$ are weak functions of centrality. The nonlinear contribution from $v_2$ is responsible for the strong centrality dependence of the correlation between $\Phi_2$ and $\Phi_{4}$ observed by the ATLAS Collaboration~\cite{Aad:2014fla} in Pb+Pb collisions. In a similar manner, the $v_5$ signal comprises a linear component proportional to $\epsilon_5$ and a leading nonlinear term involving $v_2$ and $v_3$~\cite{Gardim:2011xv,Teaney:2012gu}:
\begin{eqnarray}
\nonumber
v_5e^{i5\Phi_5} &=& a_0 \epsilon_5e^{i5\Psi_5} + a_1 \epsilon_2e^{i2\Psi_2}\epsilon_3e^{i3\Psi_3}+...\\\label{eq:0f}
              &=& c_0  e^{i5\Psi_5} + c_1 v_2v_3e^{i(2\Phi_2+3\Phi_3)}+...\;
\end{eqnarray}
This decomposition of the $v_5$ signal explains the measured EP correlation involving $\Phi_2, \Phi_3$ and $\Phi_5$~\cite{Aad:2014fla}.

Due to fluctuations of nucleon positions in the initial state, $\epsilon_n$ and $v_n$ vary from event to event, which can be described by probability distributions $p(\epsilon_n)$ and $p(v_n)$. Recent measurements by the ATLAS Collaboration~\cite{Aad:2013xma} show that the distributions $p(v_n)$ are very broad: even for events in a very narrow centrality interval, $v_2$ and $v_3$ can fluctuate from zero to several times their mean values. If events with different $v_2$ or $v_3$ values could be selected cleanly, one would be able to control directly the relative sizes of the linear and nonlinear contributions to $v_4$ and $v_5$ in Eqs.~(\ref{eq:0e}) and (\ref{eq:0f}), and hence provide an independent method of separating these two contributions. Such an event-shape selection method has been proposed in Refs.~\cite{Schukraft:2012ah,Huo:2013qma}, where events in a narrow centrality interval are further classified according to the observed ellipticity or triangularity in a forward rapidity region. These quantities are estimated from the ``flow-vector'' $q_m$ ($m=2$ and 3) as described in Section~\ref{sec:41}. This classification gives events with similar multiplicity but with very different ellipticity or triangularity. By measuring the $v_n$ and $v_m$ in a different rapidity window for each $q_m$ event class, the differential correlation between $v_m$ and $v_n$ can be obtained in an unbiased way for each centrality interval, which allows the separation of the linear and nonlinear components in $v_4$ and $v_5$. The extracted linear component of $v_4$ and $v_5$ can then be used to understand the collective response of the medium to the initial eccentricity of the same order, using an approach similar to Eq.~(\ref{eq:0dd}).

In addition to separating the linear and nonlinear effects, the correlation between $v_m$ and $v_n$ is also sensitive to any differential correlation between $\epsilon_m$ and $\epsilon_n$ in the initial state. One example is the strong anticorrelation between $\epsilon_2$ and $\epsilon_3$ predicted by the Monte Carlo (MC) Glauber model~\cite{Lacey:2013eia,Huo:2013qma}. A recent transport-model calculation shows that this correlation survives the collective expansion and appears as a similar anticorrelation between $v_2$ and $v_3$~\cite{Huo:2013qma}.

In this paper, the correlations between two flow harmonics of different order are studied using the event-shape selection method. The ellipticity or triangularity of the events is selected based on the $q_2$ or $q_3$ signal in the forward pseudorapidity range of $3.3<|\eta|<4.8$.\footnote{ATLAS uses a right-handed coordinate system with its origin at the nominal interaction point (IP) in the center of the detector and the $z$-axis along the beam pipe. The $x$-axis points from the IP to the center of the LHC ring, and the $y$-axis points upward. Cylindrical coordinates $(r,\phi)$ are used in the transverse plane, $\phi$ being the azimuthal angle around the beam pipe. The pseudorapidity is defined in terms of the polar angle $\theta$ as $\eta=-\ln\tan(\theta/2)$.} The values of $v_n$ for $n=2$ to 5 are then measured at midrapidity $|\eta|<2.5$ using a two-particle correlation method, and the correlations between two flow harmonics are obtained. The procedure for obtaining $v_n$ in this analysis is identical to that used in a previous ATLAS publication~\cite{Aad:2012bu}, which is also based on the same dataset. The main difference is that, in this analysis, the events are classified both by their centrality and by the observed $q_2$ or $q_3$ at forward pseudorapidity. Most systematic uncertainties are common to the two analyses.

\section{ATLAS detector and trigger} %\pagebreak
The ATLAS detector~\cite{Aad:2008zzm} provides nearly full solid-angle coverage of the collision point with tracking detectors, calorimeters and muon chambers. All of these are well suited for measurements of azimuthal anisotropies over a large pseudorapidity range. This analysis primarily uses two subsystems: the inner detector (ID) and the forward calorimeter (FCal). The ID is contained within the 2~T field of a superconducting solenoid magnet and measures the trajectories of charged particles in the pseudorapidity range $|\eta|<2.5$ and over the full azimuth. A charged particle passing through the ID traverses typically three modules of the silicon pixel detector (Pixel), four double-sided silicon strip modules of the semiconductor tracker (SCT), and a transition radiation tracker for $|\eta|<2$. The FCal consists of three sampling layers, longitudinal in shower depth, and covers $3.2<|\eta|< 4.9$. The energies in the FCal are reconstructed and grouped into towers with segmentation in pseudorapidity and azimuthal angle of $\Delta\eta\times\Delta\phi\approx 0.2\times0.2$. In heavy-ion collisions, the FCal is used mainly to measure the event centrality and event planes~\cite{Aad:2011yk,Aad:2012bu}. In this analysis it is also used to classify the events in terms of $q_2$ or $q_3$ in the forward rapidity region. 

The minimum-bias trigger used for this analysis requires signals in two zero-degree calorimeters (ZDC) or either of the two minimum-bias trigger scintillator (MBTS) counters. The ZDCs are positioned at $\pm$140~m from the collision point, detecting neutrons and photons with $|\eta|>8.3$, and the MBTS covers $2.1<|\eta|<3.9$ on each side of the nominal interaction point. The ZDC trigger thresholds on each side are set below the peak corresponding to a single neutron. A timing requirement based on signals from each side of the MBTS is imposed to remove beam backgrounds.

\section{Event and track selection}
\label{sec:sel}
This analysis is based on approximately 7~$\mu\mathrm{b}^{-1}$ of Pb+Pb data collected in 2010 at the LHC with a nucleon-nucleon center-of-mass energy $\sqrt{s_{_{N\!N}}}=2.76$~TeV. The offline event selection requires a reconstructed vertex and a time difference $|\Delta t| < 3$ ns between signals in the MBTS trigger counters on either side of the interaction point to suppress noncollision backgrounds. A coincidence between the ZDCs at forward and backward pseudorapidity is required to reject a variety of background processes, while maintaining high efficiency for inelastic processes. Events satisfying these conditions are further required to have a reconstructed primary vertex with $|z_{\mathrm{vtx}}|<150$~mm from the nominal center of the ATLAS detector. About 48 million events pass the requirements.

The Pb+Pb event centrality~\cite{Aad:2011yr} is characterized using the total transverse energy ($\Sigma \eT$) deposited in the FCal over the pseudorapidity range $3.2 < |\eta| < 4.9$ at the electromagnetic energy scale~\cite{Aad:2011he}. From an analysis of this distribution after all trigger and event-selection requirements, the fraction of the inelastic cross-section sampled is estimated to be $98\pm2$\%. The uncertainty associated with the centrality definition is evaluated by varying the effect of trigger and event selection inefficiencies as well as background rejection requirements in the most peripheral FCal $\Sigma \eT$ interval~\cite{Aad:2011yr}. The FCal $\Sigma \eT$ distribution is divided into a set of 5\% percentile bins. A centrality interval refers to a percentile range, starting at 0\% relative to the most central collisions. Thus the 0--5\% centrality interval corresponds to the most central 5\% of the events. An MC Glauber analysis~\cite{Miller:2007ri,Aad:2011yr} is used to estimate the average number of participating nucleons, $\npart$, for each centrality interval. These are summarized in Table~\ref{tab:cent}. Following the convention of heavy-ion analyses, the centrality dependence of the results in this paper is presented as a function of $\npart$.

\begin{table}[h!]
\centering
\caption{\label{tab:cent} The list of centrality intervals and associated values of the average number of participating nucleons $\npart$ used in this analysis. The systematic uncertainties are taken from Ref.~\cite{Aad:2011yr}.}
\begin{tabular}{lccccccc}\hline\hline
Centrality               & 0--5\%        & 5--10\%       & 10--15\%      &15--20\%   & 20--25\%      & 25--30\%      & 30--35\% \tabularnewline
$\npart$ & $382\pm2$&$330\pm3$ &$282\pm4$ &$240\pm4$& $203\pm4$ & $170\pm4$&$142\pm4$ \tabularnewline\hline
Centrality               & 35--40\%      &40--45\%       & 45--50\%      & 50--55\%   & 55--60\%     & 60--65\%      & 65--70\% \tabularnewline
$\npart$ & $117\pm4$ &$95\pm4$& $76\pm4$ &$60\pm3$  &$46\pm3$  &$35\pm3$  &$25\pm2$\tabularnewline\hline\hline
\end{tabular}
\end{table}
The harmonic flow coefficients $v_n$ are measured using tracks in the ID that are required to have transverse momentum $\pT>0.5$~GeV and $|\eta|<2.5$. At least nine hits in the silicon detectors are required for each track, with no missing Pixel hits and not more than one missing SCT hit, taking into account the effects of known dead modules. In addition, the point of closest approach of the track is required to be within 1~mm of the primary vertex in both the transverse and longitudinal directions~\cite{Aad:2011yk}. The efficiency $\epsilon(\pT,\eta)$ of the track reconstruction and track selection requirements is evaluated using simulated Pb+Pb events produced with the HIJING event generator (version 1.38b)~\cite{Gyulassy:1994ew}. The generated particles in each event are rotated in azimuthal angle according to the procedure described in Ref.~\cite{Masera:2009zz} to give harmonic flow consistent with previous ATLAS measurements~\cite{Aad:2011yk,Aad:2012bu}. The response of the detector is simulated using GEANT4~\cite{Agostinelli:2002hh,Aad:2010ah} and the resulting events are reconstructed with the same algorithms that are applied to the data. The absolute efficiency increases with $\pT$ by 7\% between 0.5~GeV and 0.8~GeV and varies only weakly for $\pT > 0.8$~GeV. However, the efficiency varies more strongly with $\eta$ and event multiplicity~\cite{Aad:2011yk}. For $\pT > 0.8$~GeV, it ranges from 72\% at $\eta \approx 0$ to 57\% for $|\eta| > 2$ in peripheral collisions, while it ranges from 72\% at $\eta \approx 0$ to about 42\% for $|\eta| > 2$ in central collisions. 

\section{Data Analysis}
\label{sec:4}

\subsection{Event-shape selection}
\label{sec:41}
The ellipticity or triangularity in each event is characterized by the so-called ``flow vector'' calculated from the transverse energy ($\eT$) deposited in the FCal~\cite{Poskanzer:1998yz,Aad:2014fla}:
\begin{eqnarray}
\label{eq:1aa}
 {\boldsymbol q}_m &=&q_me^{im\Psi^{\mathrm{obs}}_m} = \frac{\Sigma w_je^{-im \phi_j}}{\Sigma w_j}-\left\langle{\boldsymbol q}_m\right\rangle_{\mathrm{evts}},\; m =2~\mathrm{or}~3
\end{eqnarray}
where the weight $w_j$ is the $\eT$ of the $j^{\mathrm{th}}$ tower at azimuthal angle $\phi_j$ in the FCal. Subtraction of the event-averaged centroid $\left\langle{\boldsymbol q}_m\right\rangle_{\mathrm{evts}}$ in Eq.~(\ref{eq:1aa}) removes biases due to detector effects~\cite{Afanasiev:2009wq}. The angles $\Psi^{\mathrm{obs}}_{m}$ are the observed event planes, which fluctuate around the true event planes $\Phi_{m}$ due to the finite number of particles in an event. A standard technique~\cite{Barrette:1997pt} is used to remove the small residual nonuniformities in the distribution of $\Psi^{\mathrm{obs}}_m$. These procedures are identical to those used in several previous flow analyses~\cite{Afanasiev:2009wq,Aad:2012bu,Aad:2013xma,Aad:2014fla}. To reduce the detector nonuniformities at the edge of the FCal, only the FCal towers whose centroids fall within the interval $3.3<|\eta|<4.8$ are used.

\begin{figure}[!t]
\begin{center}
\includegraphics[width=1\columnwidth]{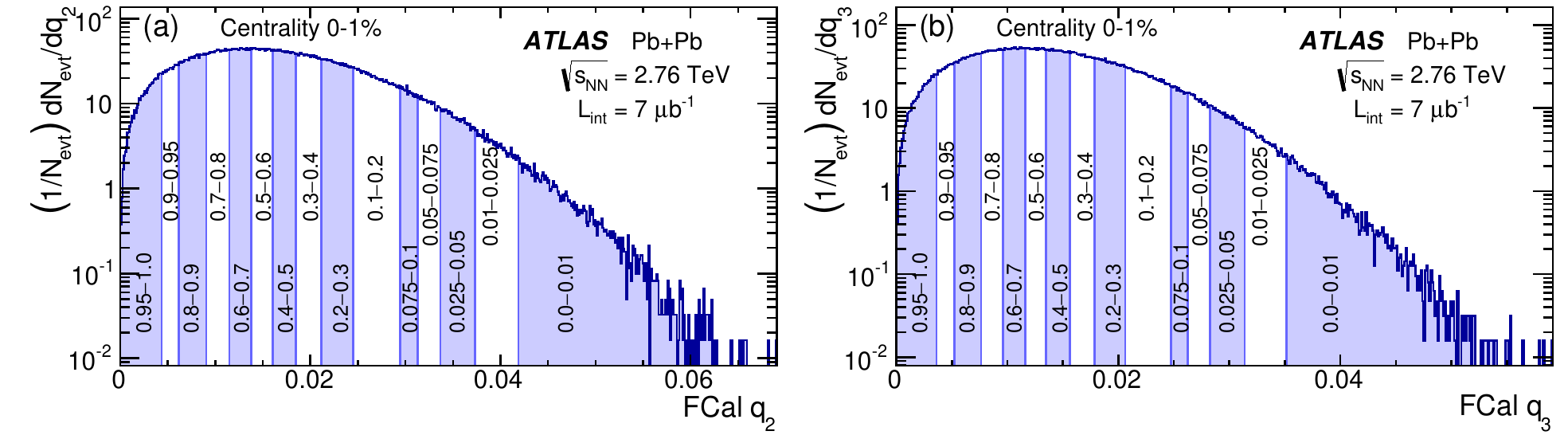}
\end{center}
\vspace*{-0.2cm}
\caption{\label{fig:0a} (Color online) The distributions of the magnitude of the flow vector, $q_2$ (left panel) and $q_3$ (right panel), calculated in the FCal via Eq.~(\ref{eq:1aa}) in the 1\% most central collisions. The vertical lines indicate the boundaries of the fifteen $q_m$ ranges, each containing a fraction of events as indicated.}
\end{figure}

%\mbox{${\boldsymbol q}_m\rightarrow \int \eT {\boldsymbol v}_m(\eT)  d\eT/\int \eT  d\eT$}. 
The ${\boldsymbol q}_m$ defined above is insensitive to the energy scale in the calorimeter. In the limit of infinite multiplicity it approaches the $\eT$-weighted single-particle flow: 
\begin{equation} 
\label{eq:1aaa}
\boldsymbol {q}_m\rightarrow \int \eT {\boldsymbol v}_m(\eT)  d\eT \left/\vphantom{ \int \eT}\right.\int \eT  d\eT\;.
\end{equation}
Hence the $q_m$ distribution is expected to follow closely the $v_m$ distribution, except that it is smeared due to the finite number of particles. Figure~\ref{fig:0a} shows the distributions of $q_2$ and $q_3$ in the 0--1\% most central collisions. These events are first divided into ten $q_m$ intervals with equal number of events. Since the intervals at the highest and lowest $q_m$ values cover much broader ranges, they are further divided into five and two smaller intervals, respectively, resulting in a total of fifteen $q_m$ intervals containing certain fractions of events. Starting at the low end of the $q_m$ distribution, there are two intervals containing a fraction 0.05 (labeled 0.95--1 and 0.9--0.95), eight intervals containing 0.1, three containing 0.025, one containing 0.015, and one containing 0.01 (this last interval spans the highest values of $q_m$). These fifteen intervals are defined separately for each 1\% centrality interval, and are then grouped together to form wider centrality intervals used in this analysis (see Table~\ref{tab:cent}). For example, the first $q_m$ interval for the 0--5\% centrality interval is the sum of the first $q_m$ interval in the five centrality intervals, 0--1\%, 1--2\%,..., 4--5\%. The default analysis uses fifteen nonoverlapping $q_m$ intervals defined in Fig.~\ref{fig:0a}. For better statistical precision, sometimes they are regrouped into wider $q_m$ intervals.
\vspace*{-0.5cm}
\subsection{Two-particle correlations}
\label{sec:42}
The two-particle correlation analysis closely follows a previous ATLAS publication~\cite{Aad:2012bu} where it is described in detail, so the analysis is only briefly summarized here. For a given event class, the two-particle correlation is measured as a function of relative azimuthal angle $\Dphi=\phi_{\rm a}-\phi_{\rm b}$ and relative pseudorapidity $\Deta=\eta_{\rm a}-\eta_{\rm b}$. The labels a and b denote the two particles in the pair, which may be selected from different $\pT$ intervals. The two-particle correlation function is constructed as the ratio of distributions for same-event pairs (or foreground pairs $S(\Delta\phi,\Delta\eta)$) and mixed-event pairs (or background pairs $B(\Delta\phi,\Delta\eta)$):
\begin{equation} 
\label{eq:1b}
C(\Delta\phi,\Delta\eta) = \frac{S(\Delta\phi,\Delta\eta)}{B(\Delta\phi,\Delta\eta)}.
\end{equation}
The mixed-event pair distribution is constructed from track pairs from two separate events with similar centrality and $z_{\mathrm{vtx}}$, such that it properly accounts for detector inefficiencies and nonuniformity, but contains no physical correlations. Charged particles measured by the ID with a pair acceptance extending up to $|\Delta\eta|=5$ are used for constructing the correlation function.

This analysis focuses mainly on the shape of the correlation function in $\Delta\phi$.  A set of 1-D $\Delta\phi$ correlation functions is built from the ratio of the foreground distributions to the background distributions, both projected onto $\Delta\phi$:
\begin{eqnarray}
\label{eq:1c}
C(\Delta\phi) = \frac{\int S(\Delta\phi,\Delta\eta) d\Delta\eta }{\int B(\Delta\phi,\Delta\eta)d\Delta\eta }.
\end{eqnarray}
The normalization is fixed by scaling the number of the mixed-event pairs to be the same as the number of same-event pairs for $2<|\Delta\eta|<5$, which is then applied to all $\Delta\eta$ slices. 

\begin{figure}[!t]
\begin{center}
\includegraphics[width=0.7\columnwidth]{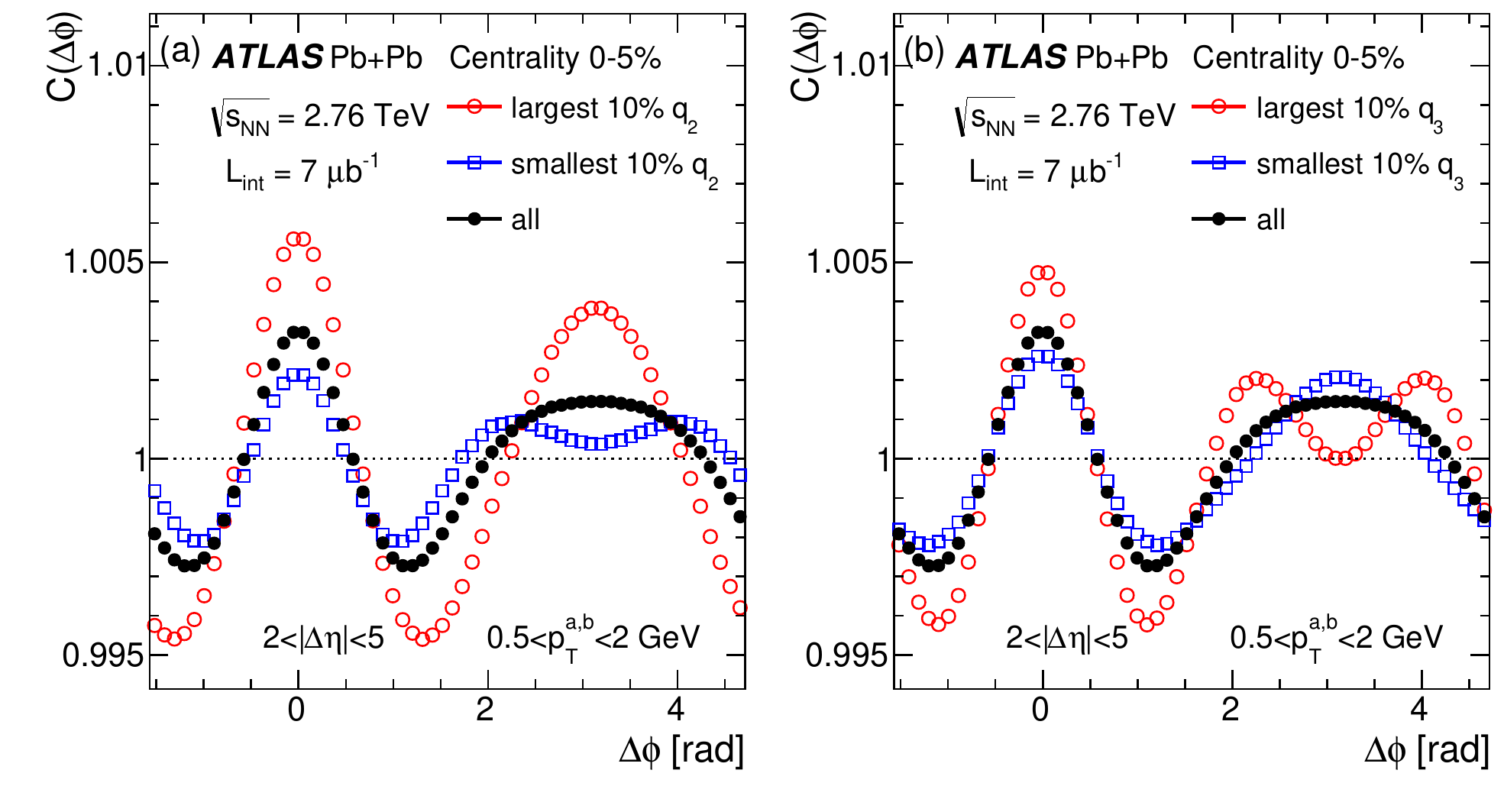}
\end{center}
\vspace*{-0.4cm}
\caption{\label{fig:0b} (Color online) The correlation functions $C(\Delta\phi)$ for pairs with $|\Delta\eta|>2$ and $0.5<\pT<2$~GeV in 0--5\% centrality. The correlation functions for events with the largest 10\% and smallest 10\% $q_m$ values are also shown for $m=2$ (left panel) and $m=3$ (right panel). The statistical uncertainties are smaller than the symbols.}
\end{figure}

Figure~\ref{fig:0b} shows the 1-D correlation functions for $2<|\Delta\eta|<5$ calculated in the low-$\pT$ region ($0.5<\ptab<2$~GeV) in the 0--5\% most central collisions. The correlation functions are also shown for events selected with the largest and smallest $q_2$ values (left panel) or $q_3$ values (right panel). The magnitude of the modulation correlates strongly with the $q_m$ value reflecting the fact that the global ellipticity or triangularity can be selected by $q_2$ or $q_3$ in the forward rapidity interval. The correlation function for events with smallest $q_2$ or largest $q_3$ values shows a double-peak structure on the away-side ($\Dphi\sim\pi$). This structure reflects the dominant contribution of the triangular flow under these $q_m$ selections. Similar double-peak structures are also observed in ultracentral Pb+Pb collisions without event-shape selection~\cite{Aad:2012bu,CMS:2013bza}.

The 1-D correlation function in $\Dphi$ is then expressed as a Fourier series:
\begin{eqnarray}
\label{eq:1d}
C(\Delta\phi) = \frac{\int C(\Delta\phi)d\Dphi}{2\pi}\left(1+2\sum_{n} v_{n,n} \cos\left(n\Delta\phi\right)\right).
\end{eqnarray}
The Fourier coefficients are calculated directly from the correlation function as $v_{n,n}= \langle\cos\left(n\Delta\phi\right)\rangle$. The single-particle azimuthal anisotropy coefficients $v_n$ are obtained via the factorization relation commonly used for collective flow in heavy ion collisions~\cite{Adcox:2002ms,Aamodt:2011by,Aad:2012bu,Chatrchyan:2013kba}:
\begin{eqnarray} 
\label{eq:1e}
v_{n,n}(\pta,\ptb)=v_n(\pta)v_n(\ptb).
\end{eqnarray}
From Eq.~(\ref{eq:1e}), $v_n$ is calculated as:
\begin{eqnarray} 
\label{eq:1f}
v_n(\pT) = v_{n,n}(\pT,\ptb)/\sqrt{v_{n,n}(\ptb,\ptb)}\,
\end{eqnarray}
where $\pta$ is simply denoted by $\pT$ from now on, and the default transverse momentum range for $\ptb$ is chosen to be $0.5<\ptb<2$~GeV, where the hydrodynamic viscous corrections are not too large. The $v_n$ values obtained using this method measure, in effect, the root-mean-square (RMS) values of the event-by-event $v_n$~\cite{Adcox:2002ms}. A detailed test of the factorization behavior was carried out~\cite{Aad:2012bu,Chatrchyan:2013kba} by comparing the $v_n(\pT)$ obtained for different $\ptb$ ranges, and factorization was found to hold to within 10\% for $\ptb<4$~GeV for the centrality ranges studied in this paper.
\subsection{Systematic uncertainties}
Other than the classification of events according to $q_m$ ($m=2$ or 3), the analysis procedure is nearly identical to the previous ATLAS measurement~\cite{Aad:2012bu} based on the same dataset. Most systematic uncertainties are the same, and they are summarized here.

The correlation function relies on the pair acceptance function to reproduce and cancel the detector acceptance effects in the foreground distribution. A natural way of quantifying the influence of detector effects on $v_{n,n}$ and $v_n$ is to express the single-particle and pair acceptance functions as Fourier series [as in Eq.~(\ref{eq:1d})], and measure the coefficients $v_{n}^{\mathrm{det}}$ and $v_{n,n}^{\mathrm{det}}$. The resulting coefficients for pair acceptance, $v_{n,n}^{\mathrm{det}}$, are the product of two single-particle acceptances $v_{n}^{\mathrm{det,a}}$ and $v_{n}^{\mathrm{det,b}}$. In general, the pair acceptance function in $\Dphi$ is quite flat: the maximum variation from its average is observed to be less than 0.001, and the corresponding $|v_{n,n}^{\mathrm{det}}|$ values are found to be less than $1.5\times 10^{-4}$. These $v_{n,n}^{\mathrm{det}}$ effects are expected to cancel to a large extent in the correlation function, and only a small fraction contributes to the uncertainties in the pair acceptance function. Three possible residual effects for $v_{n,n}^{\mathrm{det}}$ are studied in Ref.~\cite{Aad:2012bu}: 1) the time dependence of the pair acceptance, 2) the effect of imperfect centrality matching, and 3) the effect of imperfect $z_{\mathrm{vtx}}$ matching. In each case, the residual $v_{n,n}^{\mathrm{det}}$ values are evaluated by a Fourier expansion of the ratio of the pair acceptances before and after the variation. The systematic uncertainty of the pair acceptance is the sum in quadrature of these three estimates, which is $\delta v_{n,n}<5\times 10^{-6}$ for $2<|\Delta\eta|<5$. This absolute uncertainty is propagated to the uncertainty in $v_n$, and it is the dominant uncertainty when $v_n$ is small, e.g., for $v_5$ in central collisions. This uncertainty is found to be uncorrelated with the $q_m$ selection, and hence it is assumed not to cancel between different $q_m$ intervals.

A further type of systematic uncertainty includes the sensitivity of the analysis to track selection requirements and track reconstruction efficiency, variation of $v_n$ between different running periods, and trigger and event selection. The effect of the track reconstruction efficiency was evaluated in Ref.~\cite{Aad:2013xma}; the other effects were evaluated in Ref.~\cite{Aad:2012bu}. Most systematic uncertainties cancel in the correlation function when dividing the foreground distribution by the background distribution. The estimated residual effects are summarized in Table~\ref{tab:corr2}. Most of these uncertainties are expected to be correlated between different $q_m$ intervals.

\begin{table}[h]
\centering
\caption{\label{tab:corr2} Relative systematic uncertainties on the measured $v_n$ due to track selection requirements, track reconstruction efficiency, variation between different running periods, trigger selection, consistency between true and reconstructed $v_n$ in HIJING simulation, and the quadrature sum of individual terms. Most of these uncertainties are correlated between different ranges of $q_m$ ($m=2$ or 3).}
\begin{tabular}{l| c c  c c c c |c}\hline\hline
                                       &$\textrm{      }$& $v_2$   & $v_3$   & $v_4$   & $v_5$&$\textrm{      }$   &$q_m$-dependent\\\hline
Track selection [\%]                   && 0.3     & 0.3     & 1.0     & 2.0     &&yes\\\hline
Track reconstruction efficiency [\%]                &&0.1--1.0  & 0.2--1.5 & 0.2--2.0 & 0.3--2.5 &&yes\\\hline
Running periods [\%]                    &&0.3--1.0  & 0.7--2.1 & 1.2--3.1 & 2.3     &&no \\\hline
Trigger [\%]                           &&0.5--1.0  & 0.5--1.0 & 0.5--1   & 1.0     &&yes\\\hline
MC closure and occupancy effects [\%]   && 1.0     & 1.5     & 2.0     & 3.5     &&yes\\\hline
Sum of above [\%]                       &&1.2--2.0  & 1.8--3.2 &2.6--4.4  & 4.7--5.4 &&\\\hline\hline
\end{tabular}
\end{table}
Finally, due to the anisotropy of particle emission, the detector occupancy is expected to be larger in the direction of the event plane, where the particle density is larger. Any occupancy effects depending on azimuthal angle may lead to a small angle-dependent efficiency variation, which may slightly reduce the measured $v_n$ coefficients. The magnitude of such an occupancy-dependent variation in tracking efficiency is evaluated using the HIJING simulation with flow imposed on the generated particles~\cite{Aad:2013xma}. The reconstructed $v_n$ values are compared to the generated $v_n$ signal. The differences are taken as an estimate of the systematic uncertainties. These differences are found to be a few percent or less, and are included in Table~\ref{tab:corr2}. Since this effect is proportional to the flow signal, it is expected to partially cancel between different $q_m$ ranges.

\section{Results}
\label{sec:6}
\subsection{Fourier coefficients $v_n$ and their correlations with $q_m$}
\label{sec:res0} 
\begin{figure}[!t]
\begin{center}
\includegraphics[width=0.39\columnwidth]{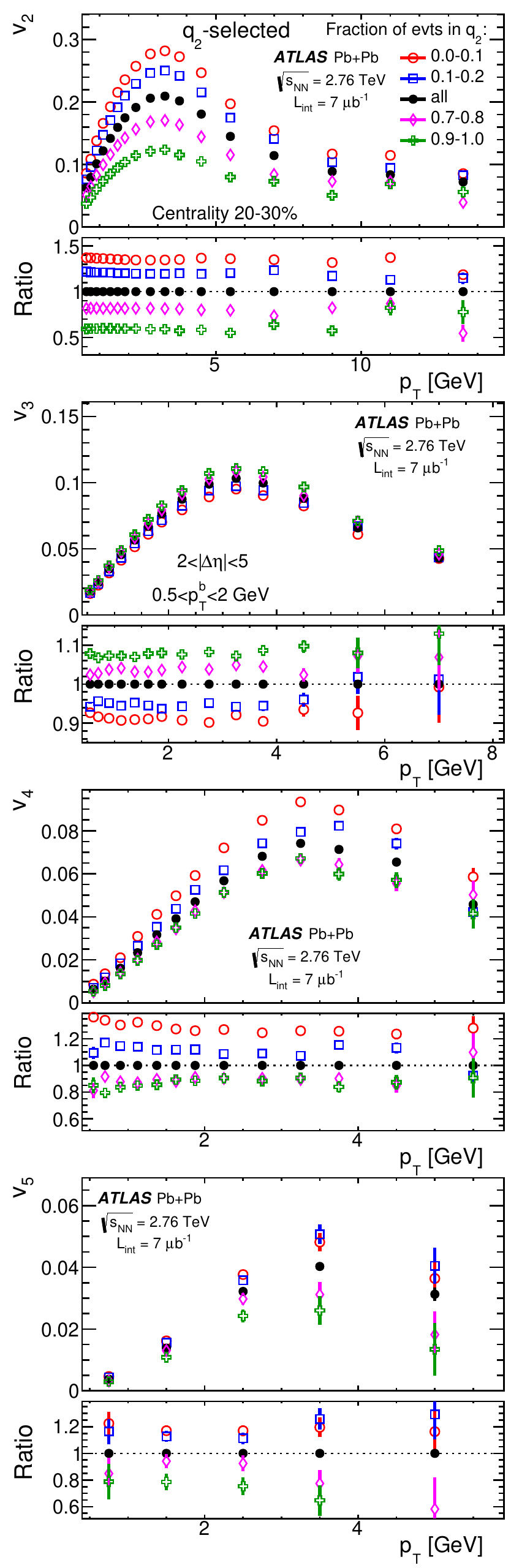}\hspace*{0.1\columnwidth}\includegraphics[width=0.39\columnwidth]{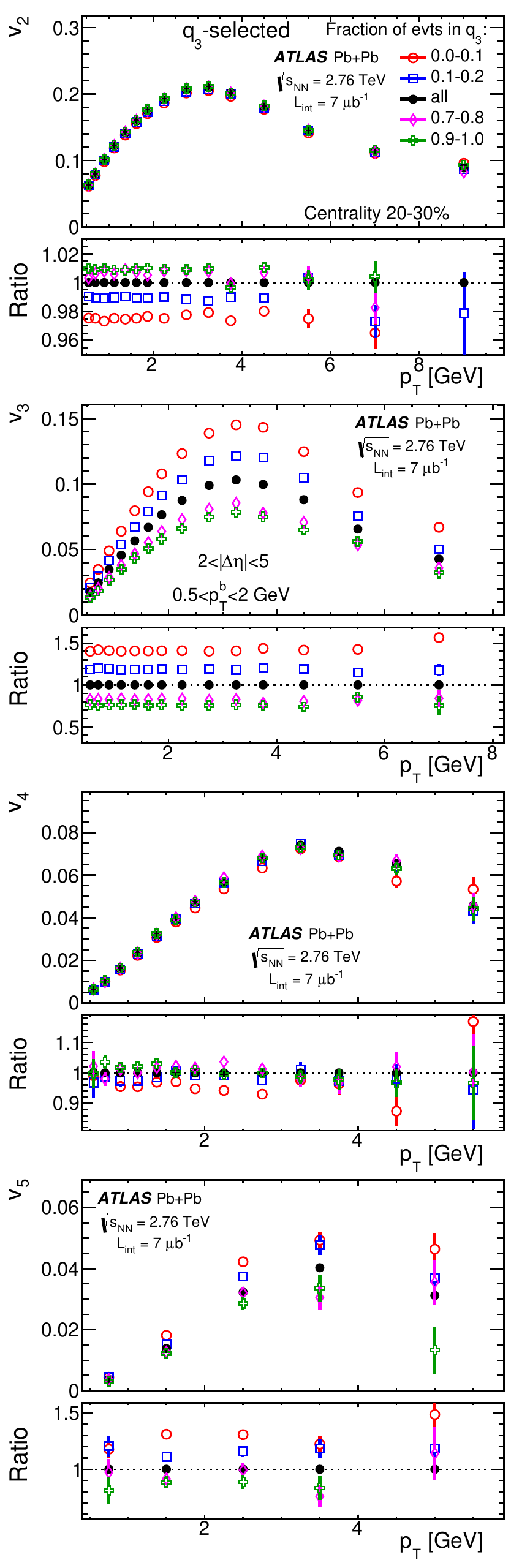}
\end{center}
\caption{\label{fig:0c} (Color online) The harmonic flow coefficients $v_n(\pT)$ in the 20--30\% centrality interval for events selected on either $q_2$ (left column) or $q_3$ (right column) for $n=2$ (top row), $n=3$ (second row), $n=4$ (third row) and $n=5$ (bottom row). They are calculated for reference $\pT$ of $0.5<\ptb <2$~GeV [Eq.~(\ref{eq:1f})]. The top part of each panel shows the $v_n(\pT)$ for events in the 0--0.1, 0.1--0.2, 0.7--0.8 and 0.9--1 fractional ranges of $q_m$ (open symbols) as well as for inclusive events without $q_m$ selection (solid symbols). The bottom part of each panel shows the ratios of the $v_n(\pT)$ for $q_m$-selected events to those obtained for all events. Only statistical uncertainties are shown.}
\end{figure}

Figure~\ref{fig:0c} shows the $v_n(\pT)$ for $n=2$ to 5 extracted via Eq.~(\ref{eq:1f}) for events in the 20--30\% centrality interval. The results show nontrivial correlations with both the $q_2$ (left column) or $q_3$ (right column) selections. In the case of the $q_2$ selection, the $v_2$ values are largest for events selected with the largest $q_2$, and smallest for events selected with the smallest $q_2$, with a total change of more than a factor of two. A similar dependence on $q_2$ is also seen for $v_4(\pT)$ and $v_5(\pT)$ (two bottom panels). In contrast, the extracted $v_3(\pT)$ values are anticorrelated with $q_2$; the overall change in $v_3(\pT)$ is also significantly smaller ($<20\%$ across the $q_2$ range). In the case of the $q_3$ selection, a strong positive correlation is observed for $v_3$ and $v_5$, and a weak anticorrelation is observed for $v_2$ and $v_4$. All these correlations are observed to be nearly independent of $\pT$, suggesting that the response of $v_n$ to the change in the event shape is largely independent of $\pT$. As a consistency check, the inclusive results without $q_m$ selection are compared with previously published results from Ref.~\cite{Aad:2012bu}: the differences are less than $0.6$\% for $v_2$ and increase to 2--3\% for higher harmonics, which are well within the systematic uncertainties quoted in Table~\ref{tab:corr2}.

\begin{figure}[!t]
\begin{center}
\includegraphics[width=0.45\columnwidth]{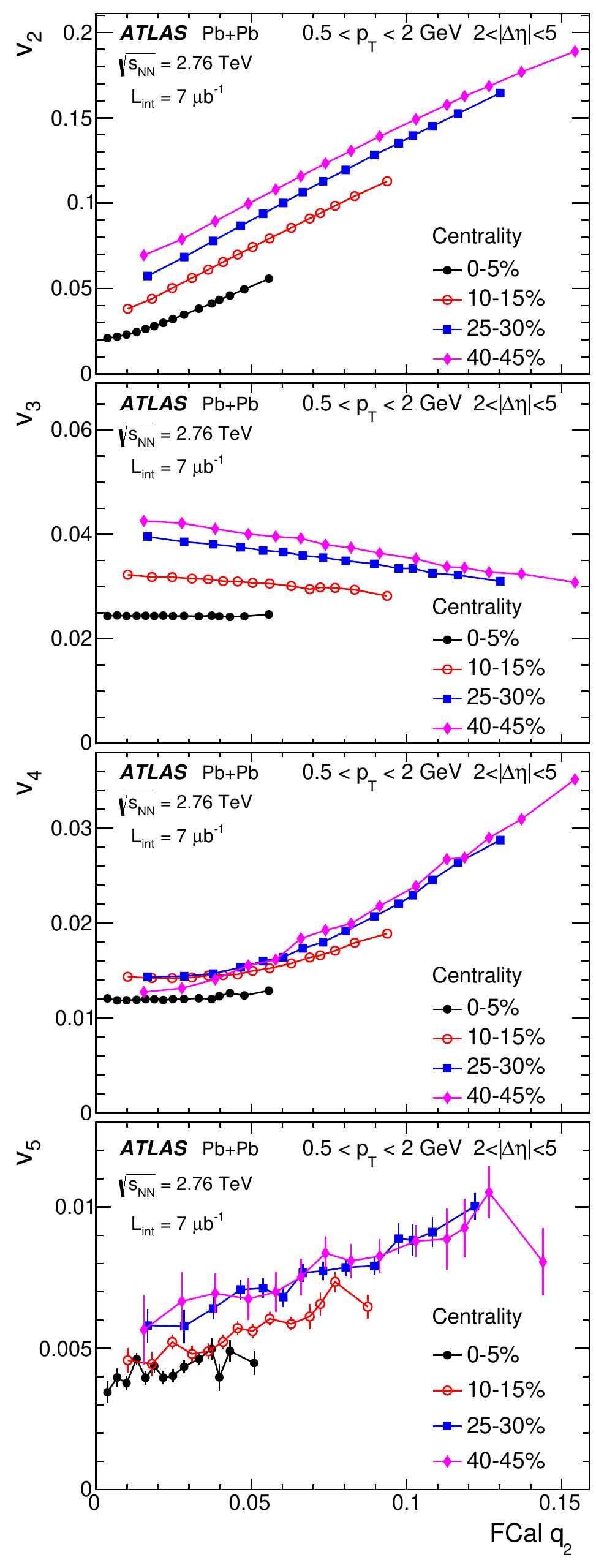}\hspace*{0.1\columnwidth}\includegraphics[width=0.45\columnwidth]{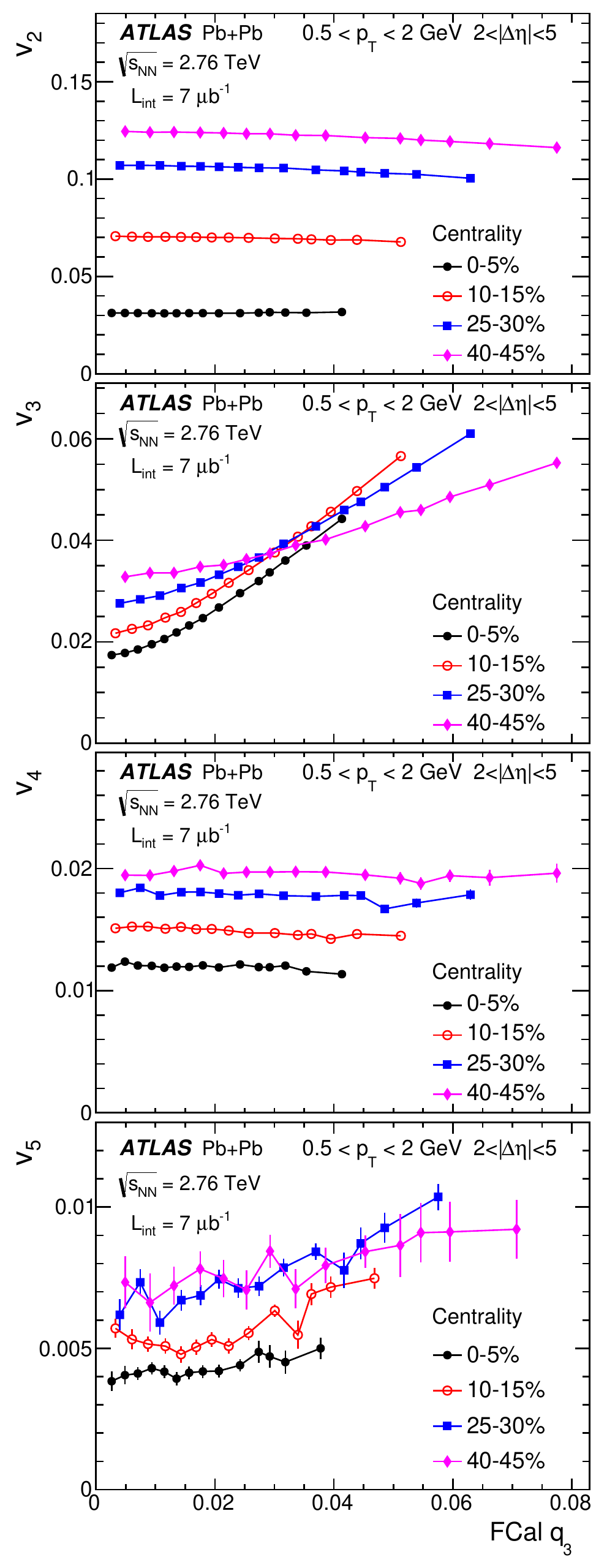}
\end{center}
\vspace*{-0.4cm}
\caption{\label{fig:0cc} (Color online) The correlations between $v_n$ and $q_2$ (left column) and $q_3$ (right column) in four centrality intervals with $n=2$ (top row), $n=3$ (second row), $n=4$ (third row) and $n=5$ (bottom row), where $v_n$ is calculated in $0.5<\pT<2$~GeV. Only statistical uncertainties are shown. The lines connecting data points are for guidance only.}
\end{figure}

Figure~\ref{fig:0cc} shows the correlation between $v_n$ and $q_m$ for $m=2$ (left column) and $m=3$ (right column) in several centrality intervals in a low $\pT$ range ($0.5<\pT<2$~GeV). Since the $v_n$--$q_m$ correlation depends only weakly on $\pT$, this plot captures the essential features of the correlation between $v_n$ and $q_m$ shown in Fig.~\ref{fig:0c}. Due to the finite number of particles in an event, the measured $q_m$ values fluctuate relative to the true values, diluting the correlations with $v_n$. The influence of smearing on the $q_2$ is much smaller than that for the $q_3$ simply because the $v_2$ signal is much bigger than the $v_3$ signal. But since both the $v_m$--$q_m$ and $v_n$--$q_m$ correlations are measured, the results are presented directly as $v_m$--$v_n$ correlations for various $q_m$ selections. The level of detail contained in the $v_m$--$v_n$ correlation is controlled by the dynamic range of $v_m$ when varying the $q_m$ selection. This dynamic range depends strongly on event centrality. For example, in the 10--15\% centrality interval, $v_2$ is varied by a factor of 3.1 by selecting on $q_2$ and $v_3$ is varied by a factor of 2.4 by selecting on $q_3$. In the 40--45\% centrality interval, however, due to stronger statistical smearing of $q_m$, the $v_2$ and $v_3$ are only varied by a factor of 2.7 and 1.7, respectively. Hence the event-shape selection is precise in central and midcentral collisions and is expected to be less precise in peripheral collisions.

In general, correlations $v_m$--$q_m$ and $v_n$--$q_m$ can be measured in different $\pT$ ranges, and the derived $v_m$--$v_n$ correlation can be categorized into three types: 1) the correlation between $v_m$ in two different $\pT$ ranges, $v_m\{\pta\}$--$v_m\{\ptb\}$, 2) the correlation between $v_m$ and another flow harmonic of different order $v_n$ in the same $\pT$ range, $v_m\{\pT \}$--$v_n\{\pT \}$, and 3) the correlation between $v_m$ and $v_n$ in different $\pT$ ranges, $v_m\{\pta\}$--$v_n\{\ptb\}$. However, the $v_m\{\pta\}$--$v_n\{\ptb\}$ correlation can be obtained by combining two correlations $v_m\{\pta\}$--$v_m\{\ptb \}$ and $v_m\{\ptb\}$--$v_n\{\ptb\}$, so it does not carry independent information. This paper therefore focuses on the first two types of correlation.

The results for $v_m$--$v_n$ correlations are organized as follows. Section~\ref{sec:res1} presents correlations of $v_2$ or $v_3$ between two different $\pT$ ranges. The $v_2$--$v_3$ correlations are discussed in Sec.~\ref{sec:res2}. This is followed by $v_2$--$v_4$ and $v_3$--$v_4$ correlations in Sec.~\ref{sec:res3} and $v_2$--$v_5$ and $v_3$--$v_5$ correlations in Sec.~\ref{sec:res4}, where a detailed analysis is performed to separate the linear and nonlinear components of $v_4$ and $v_5$. The eccentricity scaling behavior of the extracted linear component of $v_n$ is presented in Sec.~\ref{sec:res5}.

\subsection{Correlation of $v_2$ or $v_3$ between two different $\pT$ ranges}
\label{sec:res1} 
\begin{figure}[!h]
\begin{center}
\includegraphics[width=0.8\columnwidth]{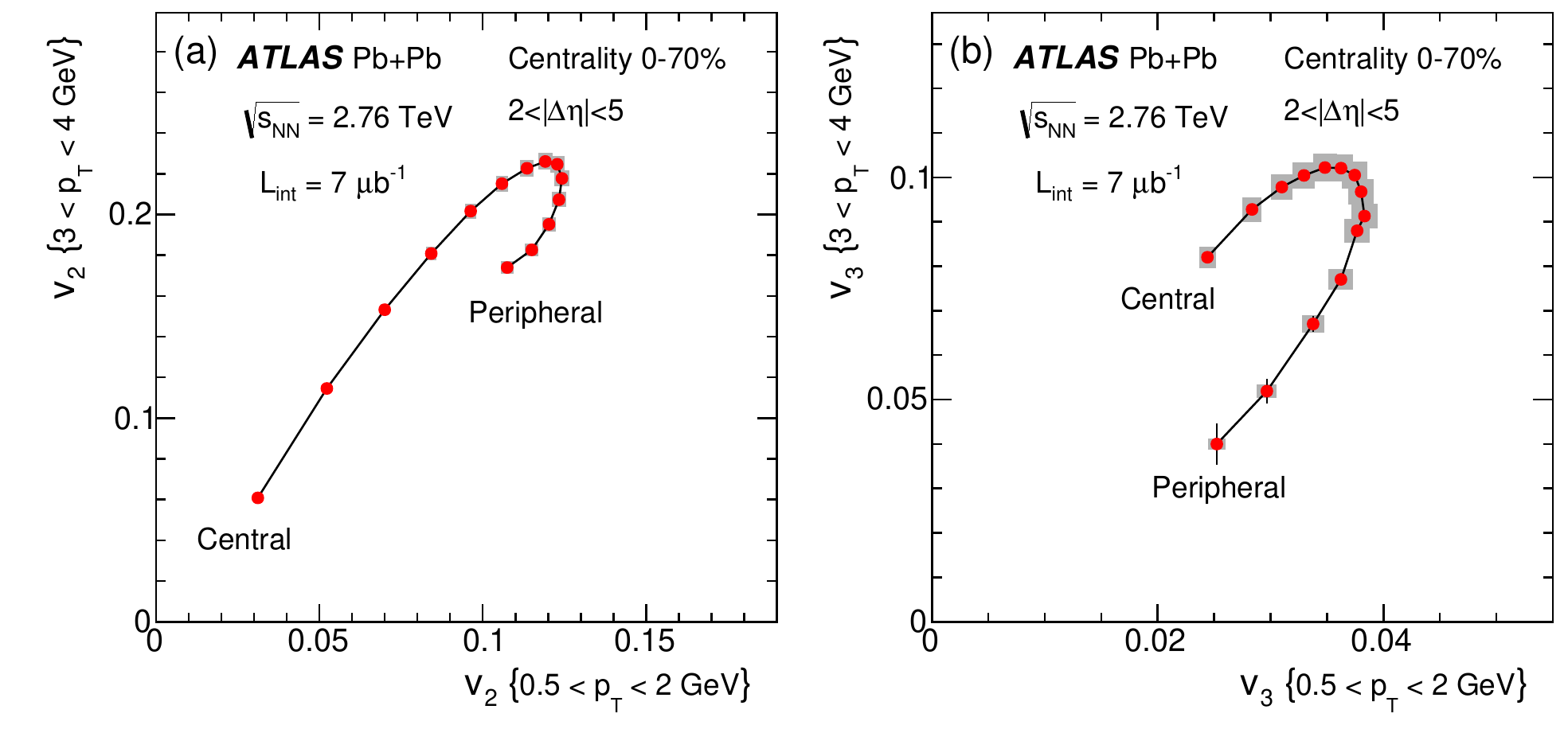}
\end{center}
\caption{\label{fig:1} (Color online)  The correlation of the $v_m$ between $0.5<\pT<2$~GeV ($x$-axis) and $3<\pT<4$~GeV ($y$-axis) for $m=2$ (left panel) and $m=3$ (right panel). The $v_m$ values are calculated for fourteen 5\% centrality intervals in the centrality range 0--70\% without event-shape selection. The data points are connected to show the boomerang trend from central to peripheral collisions as indicated. The error bars and shaded boxes represent the statistical and systematic uncertainties, respectively. These uncertainties are often smaller than the symbol size.}
\end{figure}

Figure~\ref{fig:1} shows the correlation of $v_m$ for $m=2$ (left panel) or $m=3$ (right panel) between two $\pT$ ranges for various centrality intervals. The $x$-axis represents $v_m$ values in the $0.5<\pT<2$~GeV range, while the $y$-axis represents $v_m$ values from a higher range of $3<\pT<4$~GeV. Each data point corresponds to a 5\% centrality interval within the overall centrality range of 0--70\%. Going from central collisions (left end of the data points) to the peripheral collisions (right end of the data points), $v_m$ first increases and then decreases along both axes, reflecting the characteristic centrality dependence of $v_m$, well known from previous flow analyses~\cite{ALICE:2011ab,Aad:2012bu}.  The rate of decrease is larger at higher $\pT$, resulting in a ``boomerang-like'' structure in the correlation. The stronger centrality dependence of $v_m$ at higher $\pT$ is consistent with larger viscous-damping effects expected from hydrodynamic calculations~\cite{Luzum:2008cw}.
\begin{figure}[!t]
\begin{center}
\includegraphics[width=0.92\columnwidth]{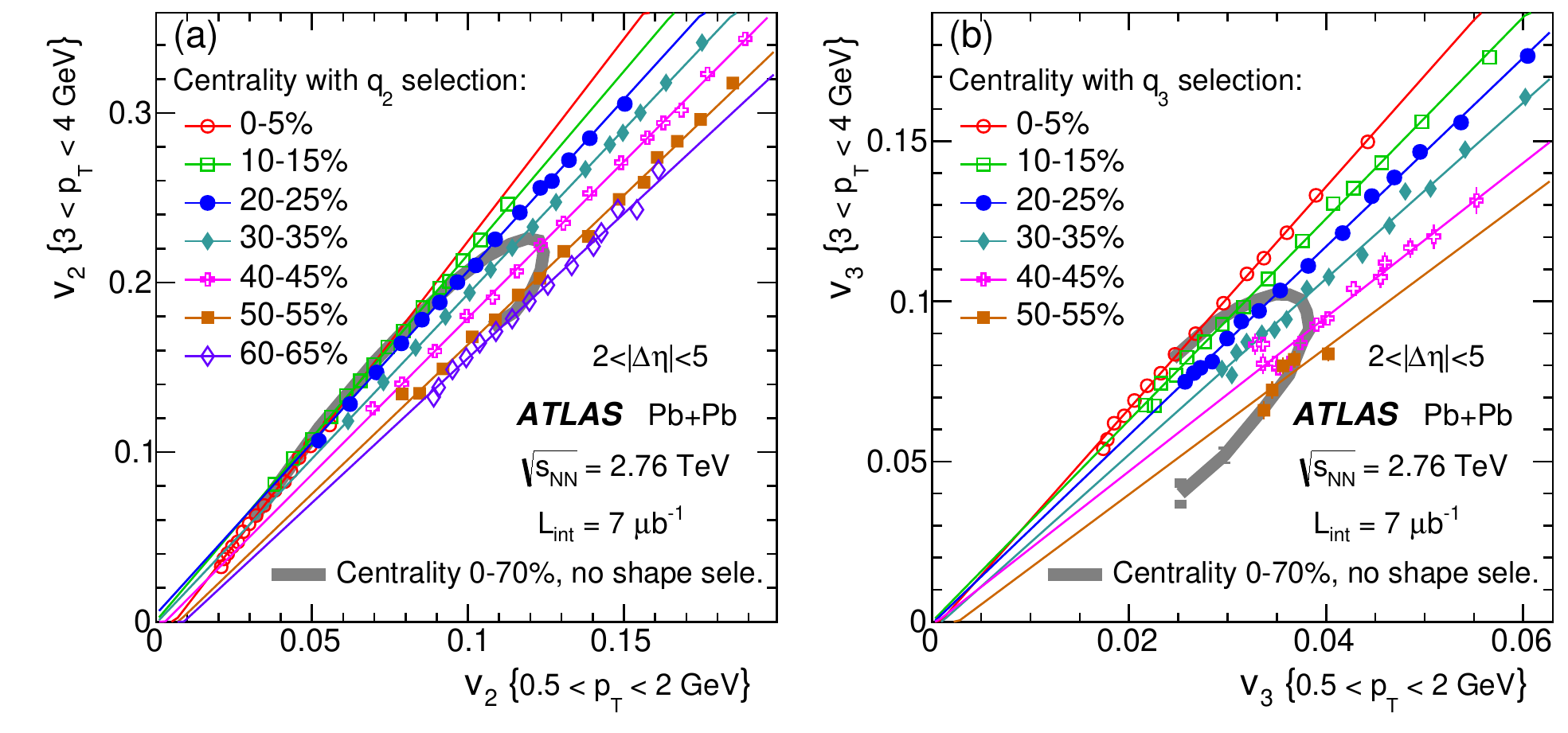}
\end{center}
\caption{\label{fig:1b} (Color online) The correlation of $v_m$ between the $0.5<\pT<2$~GeV ($x$-axis) and $3<\pT<4$~GeV range ($y$-axis) for $m=2$ (left panel) and $m=3$ (right panel) in various centrality intervals. The data points are calculated in various $q_m$ intervals defined in Fig.~\ref{fig:0a} for each centrality, and they increase monotonically with increasing $q_m$ value. These data are overlaid with the centrality dependence without $q_m$ selection from Fig.~\ref{fig:1}. The thin solid straight lines represent a linear fit of the data in each centrality interval, and error bars represent the statistical uncertainties.}
\end{figure}

In the next step, events in each centrality interval are further divided into $q_m$ intervals, as described in Sec.~\ref{sec:41}. With this further subdivision, each data point in Fig.~\ref{fig:1} turns into a group of data points, which may follow a different correlation pattern. These data points are shown in Fig.~\ref{fig:1b} (markers) overlaid with the overall centrality dependence prior to the event-shape selection from Fig.~\ref{fig:1} (the ``boomerang''). For clarity, the results are shown only for seven selected centrality intervals. Unlike the centrality dependence, the $v_m$ correlation within a given centrality interval approximately follows a straight line passing very close to the origin. The small nonzero intercepts can be attributed to a residual centrality dependence of the $v_m$--$v_m$ correlation within the finite centrality intervals used. This approximately linear correlation suggests that, once the event centrality or the overall event multiplicity is fixed, the viscous-damping effects on $v_m$ change very little with the variation of the event shape via $q_m$ selection. The influence of viscosity on flow harmonics is mainly controlled by the event centrality (or the overall system size).
\subsection{$v_2$--$v_3$ correlation}
\label{sec:res2} 

\begin{figure}[!h]
\begin{center}
\includegraphics[width=0.93\columnwidth]{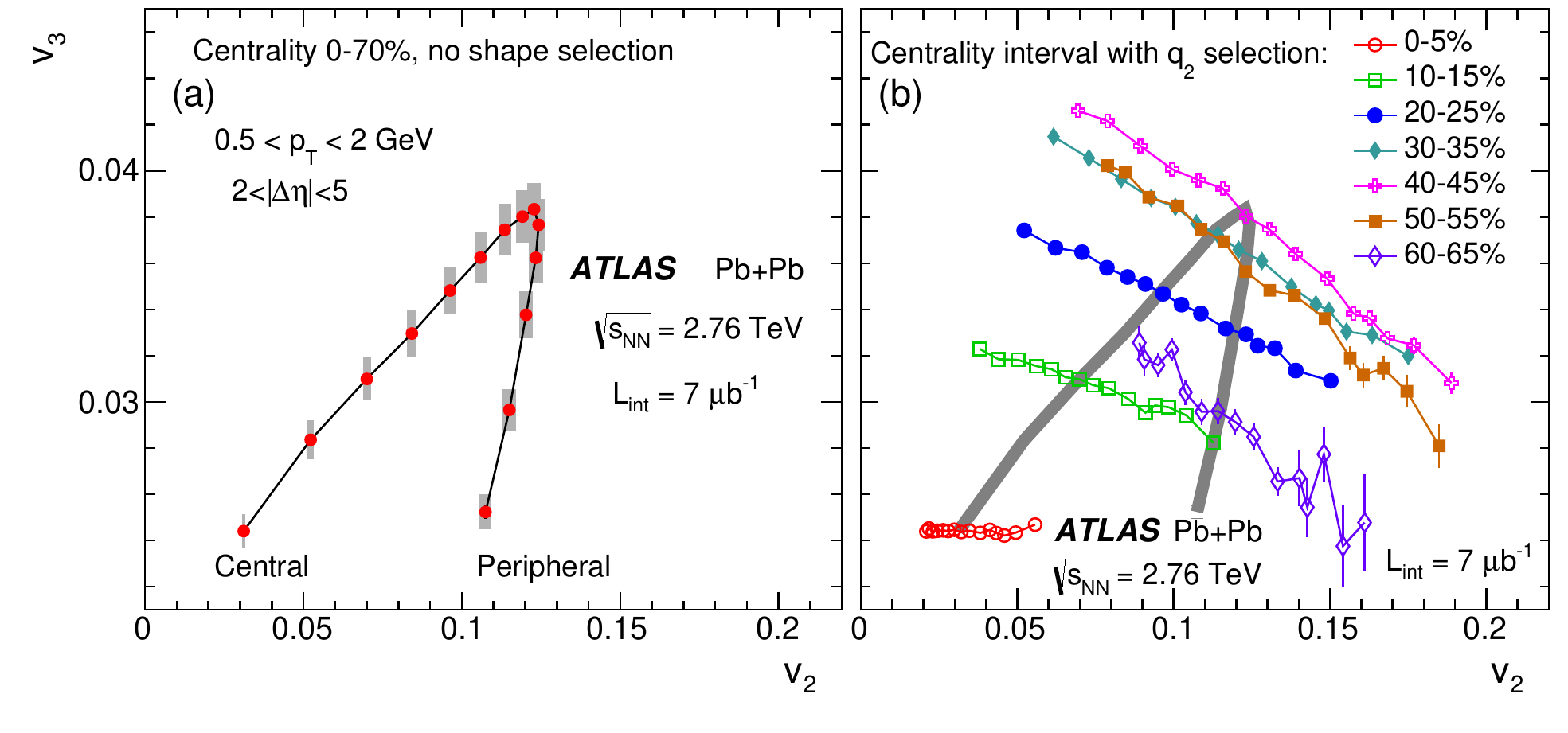}
\end{center}
\caption{\label{fig:2} (Color online) The correlation of $v_2$ ($x$-axis) with $v_3$ ($y$-axis) both measured in $0.5<\pT<2$ GeV. The left panel shows the $v_2$ and $v_3$ values for fourteen 5\% centrality intervals over the centrality range 0--70\% without event-shape selection. The data points are connected to show the boomerang trend from central to peripheral collisions as indicated. The right panel shows the $v_2$ and $v_3$ values in the fifteen $q_2$ intervals in seven centrality ranges (markers) with larger $v_2$ value corresponding to larger $q_2$ value; they are overlaid with the centrality dependence from the left panel.  The error bars and shaded boxes represent the statistical and systematic uncertainties, respectively.}
\end{figure}

Figure~\ref{fig:2}(a) shows the centrality dependence of the correlation between $v_2$ and $v_3$ measured in $0.5<\pT<2$~GeV. The boomerang-like structure in this case reflects mostly the fact that $v_3$ has a much weaker centrality dependence than $v_2$~\cite{Aad:2012bu}. Figure~\ref{fig:2}(b) overlays the centrality dependence of the $v_2$--$v_3$ correlation (thick solid line) with those obtained for different $q_2$ event classes (markers). The correlation within a fixed centrality interval follows a path very different from the centrality dependence: the $v_2$ and $v_3$ are always anticorrelated with each other within a given centrality, whereas they are positively correlated as a function of centrality. Since the $v_2$ and $v_3$ are driven by the initial eccentricities, $v_2\propto \epsilon_2$ and $v_3\propto \epsilon_3$, one may expect similar anticorrelation between $\epsilon_2$ and $\epsilon_3$. Indeed, a calculation based on a multiphase transport model~\cite{Lin:2004en} shows that such anticorrelations exist in the initial geometry and they are transferred into similar anticorrelations between $v_2$ and $v_3$ by the collective expansion~\cite{Huo:2013qma}.

\begin{figure}[t]
\begin{center}
\includegraphics[width=0.95\columnwidth]{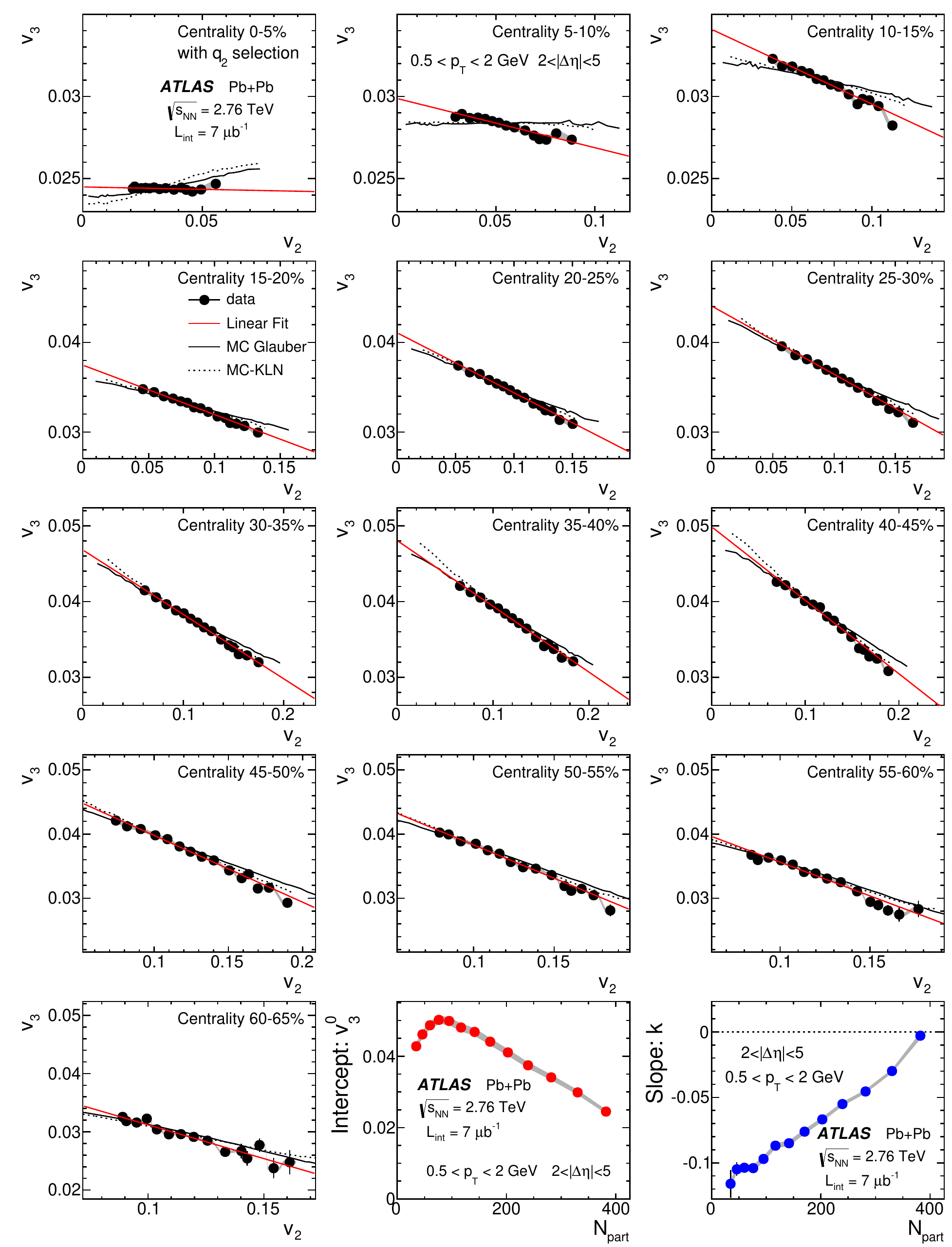}
\end{center}
\caption{\label{fig:2c} (Color online) The correlation of $v_2$ ($x$-axis) with $v_3$ ($y$-axis) in $0.5<\pT<2$~GeV for fifteen $q_2$ selections in thirteen 5\% centrality intervals. The data are compared with the rescaled $\epsilon_2$--$\epsilon_3$ correlation from MC Glauber and MC-KLN models in the same centrality interval. The data are also parameterized with a linear function [Eq.~(\ref{eq:2a})], taking into account both the statistical and systematic uncertainties. The $\npart$ dependence of the fit parameters is shown in the last two panels. The error bars and shaded bands represent the statistical and systematic uncertainties, respectively.}
\end{figure}
In order to illustrate this anticorrelation more clearly, the $v_2$--$v_3$ correlation data are replotted in Fig.~\ref{fig:2c}, separately for each centrality. The data are compared with the $\epsilon_2$--$\epsilon_3$ correlations calculated via Eq.~(\ref{eq:0c}) from the MC Glauber model~\cite{Miller:2007ri} and the MC-KLN model~\cite{Drescher:2006pi}. The MC-KLN model is based on the MC Glauber model, but takes into account gluon saturation effects in the initial geometry. One hundred million events were generated for each model and grouped into centrality intervals according to the impact parameter. The RMS $\epsilon_n$ value for each centrality interval is rescaled by a factor $s_n$ to match the inclusive $v_n$ value, which effectively is also the RMS value of $v_n$ [see Eq.~(\ref{eq:1f})]: 
\begin{eqnarray}
\label{eq:2aa}
s_n= \frac{v_n}{\sqrt{\langle\epsilon_n^2\rangle}}\;.
\end{eqnarray}
The parameter $s_n$ changes with centrality but is assumed to be a constant within a given centrality interval. These constants are then used to rescale the $\epsilon_2$--$\epsilon_3$ correlation to be compared with the $v_2$--$v_3$ correlation in each centrality interval, as shown in Fig.~\ref{fig:2c}.
In most centrality intervals the rescaled $\epsilon_2$--$\epsilon_3$ correlation shows very good agreement with the $v_2$--$v_3$ correlation seen in the data. However, significant deviations are observed in more central collisions (0--20\% centrality range). Therefore, the $v_2$--$v_3$ correlation data presented in this analysis can provide valuable constraints for further tuning of the initial geometry models. The $v_2$--$v_3$ correlations in Fig.~\ref{fig:2c} are parameterized by a linear function:
\begin{eqnarray}
\label{eq:2a}
v_3 = kv_2 + v_3^0\;,
\end{eqnarray}
where the intercept $v_3^0$ provides an estimate of the asymptotic $v_3$ value for events that have zero $v_2$ for each centrality. The fit parameters are summarized as a function of centrality ($\npart$) in the last two panels of Fig.~\ref{fig:2c}. 

\subsection{$v_2$--$v_4$ and $v_3$--$v_4$ correlations}
\label{sec:res3} 
\begin{figure}[!h]
\begin{center}
\includegraphics[width=0.93\columnwidth]{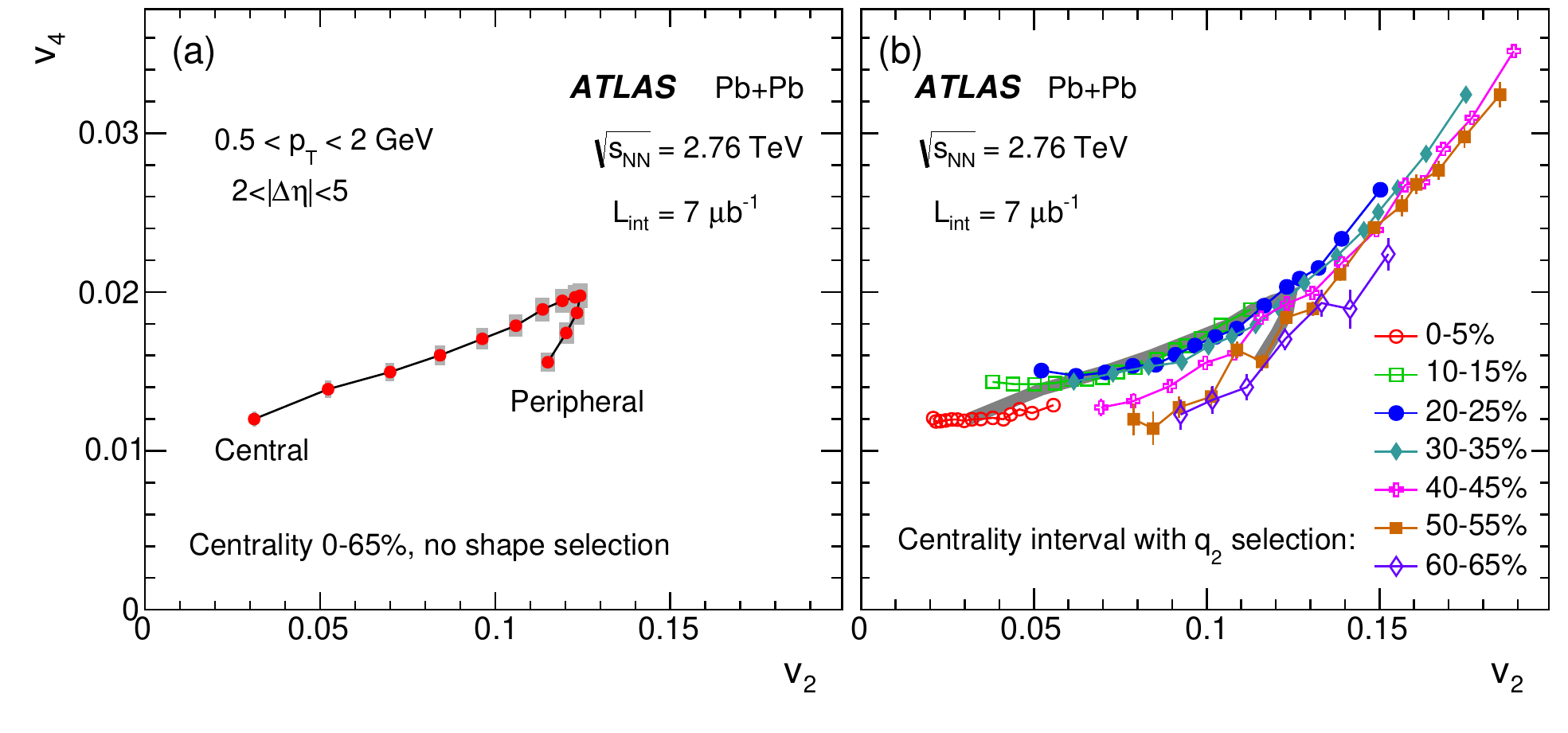}
\end{center}
\caption{\label{fig:3}  (Color online) The correlation of $v_2$ ($x$-axis) with $v_4$ ($y$-axis) both measured in $0.5<\pT<2$ GeV. The left panel shows the $v_2$ and $v_4$ values for thirteen 5\% centrality intervals over the centrality range 0--65\% without event-shape selection. The data points are connected to show the boomerang trend from central to peripheral collisions as indicated. The right panel shows the $v_2$ and $v_4$ values in different $q_2$ intervals in seven centrality ranges (markers) with larger $v_2$ value corresponding to larger $q_2$ value; they are overlaid with the centrality dependence from the left panel.  The error bars and shaded boxes represent the statistical and systematic uncertainties, respectively.}
\end{figure}
Figure~\ref{fig:3}(a) shows the correlation between $v_2$ and $v_4$ in $0.5<\pT<2$~GeV prior to the event-shape selection. The boomerang-like structure is less pronounced than that for the $v_2$--$v_3$ correlation shown in Fig.~\ref{fig:2}(a). Figure~\ref{fig:3}(b) shows the $v_2$--$v_4$ correlation for different $q_2$ event classes (markers) overlaid with the centrality dependence taken from Fig.~\ref{fig:3}(a) (thick solid line). The correlation within a given centrality interval is broadly similar to the trend of the correlation without event-shape selection, but without any boomerang effect. Instead, the shape of the correlation exhibits a nonlinear rise for large $v_2$ values.
\begin{figure}[!h]
\begin{center}
\includegraphics[width=0.9\columnwidth]{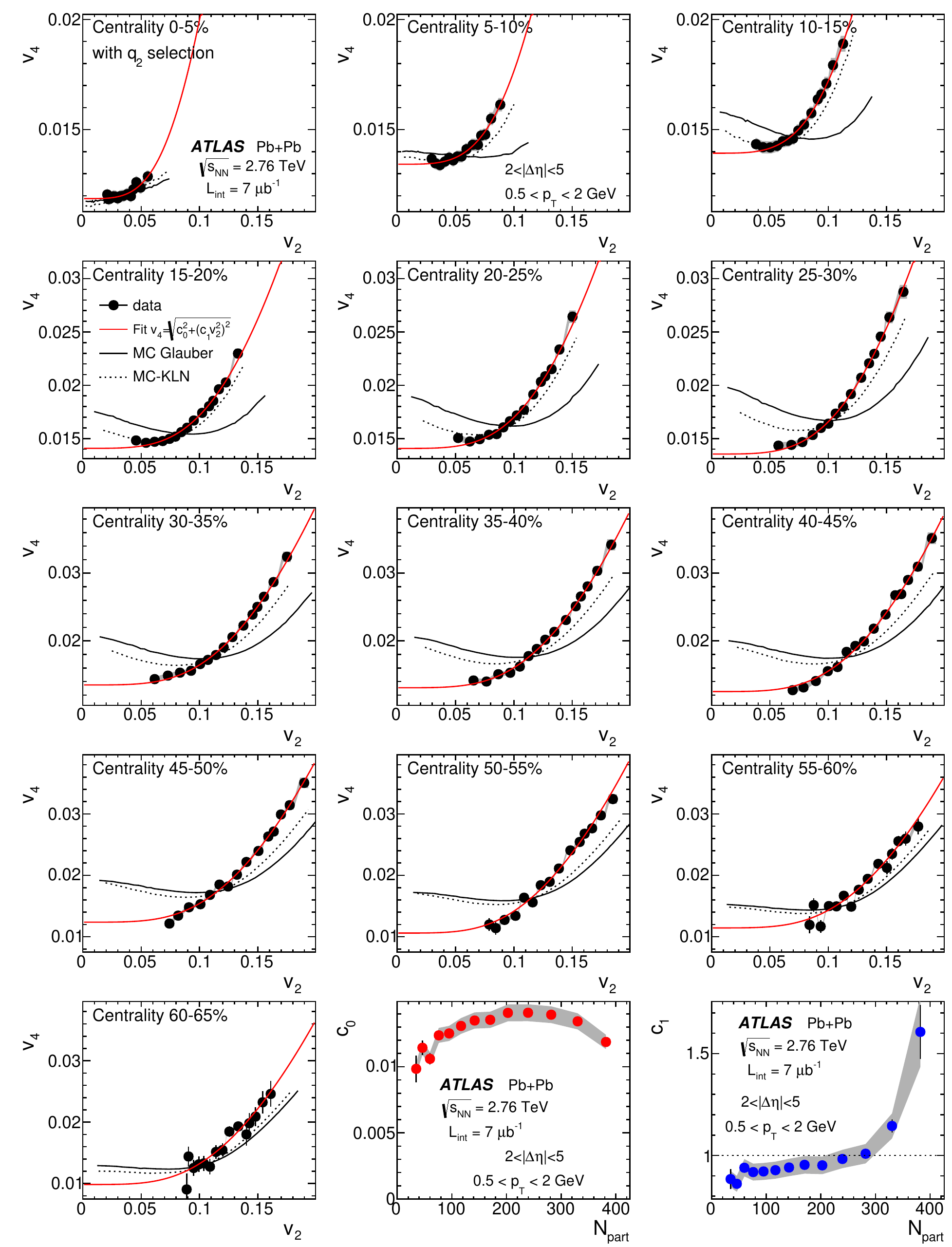}
\end{center}
\caption{\label{fig:3c} (Color online) The correlation of $v_2$ ($x$-axis) with $v_4$ ($y$-axis) in $0.5<\pT<2$~GeV for fifteen $q_2$ selections in thirteen 5\% centrality intervals. The data are compared with the rescaled $\epsilon_2$--$\epsilon_4$ correlation from MC Glauber and MC-KLN models in the same centrality interval. The data are also parameterized with Eq.~(\ref{eq:2b}), taking into account both statistical and systematic uncertainties. The $\npart$ dependence of the fit parameters is shown in the last two panels.  The error bars and shaded bands represent the statistical and systematic uncertainties, respectively.}
\end{figure}

To understand further the role of the linear and nonlinear contributions to $v_4$, the $v_2$--$v_4$ correlation data in Fig.~\ref{fig:3} are shown again in Fig.~\ref{fig:3c}, separately for each centrality. The data are compared with the $\epsilon_2$--$\epsilon_4$ correlation rescaled according to Eq.~(\ref{eq:2aa}). The rescaled $\epsilon_2$--$\epsilon_4$ correlations fail to describe the data, suggesting that the linear component alone associated with $\epsilon_4$ in Eq.~(\ref{eq:0e}) is not sufficient to explain the measured $v_2$--$v_4$ correlation.

To separate the linear and nonlinear components in the $v_2$--$v_4$ correlation, the data are fitted to the following functional form:
\begin{eqnarray}
\label{eq:2b}
v_4&=& \sqrt{c_{0}^2 + (c_{1}v_2^2)^{2}}\;.
\end{eqnarray}
This function is derived from Eq.~(\ref{eq:0e}), by ignoring the higher-order nonlinear terms (those in ``...'') and a possible cross-term that is proportional to $\langle\cos4(\Psi_2-\Psi_4)\rangle$. The fits, which are shown in Fig.~\ref{fig:3c}, describe the data well for all centrality intervals. The excellent description of the data by the fits suggests that either the contributions from higher-order nonlinear terms and $\langle\cos4(\Psi_2-\Psi_4)\rangle$ are small, or the cross-term is, in effect, included in the nonlinear component of the fits. The centrality ($\npart$) dependence of the fit parameters is shown in the last two panels of Fig.~\ref{fig:3c}. 

\begin{figure}[!t]
\begin{center}
\includegraphics[width=0.8\columnwidth]{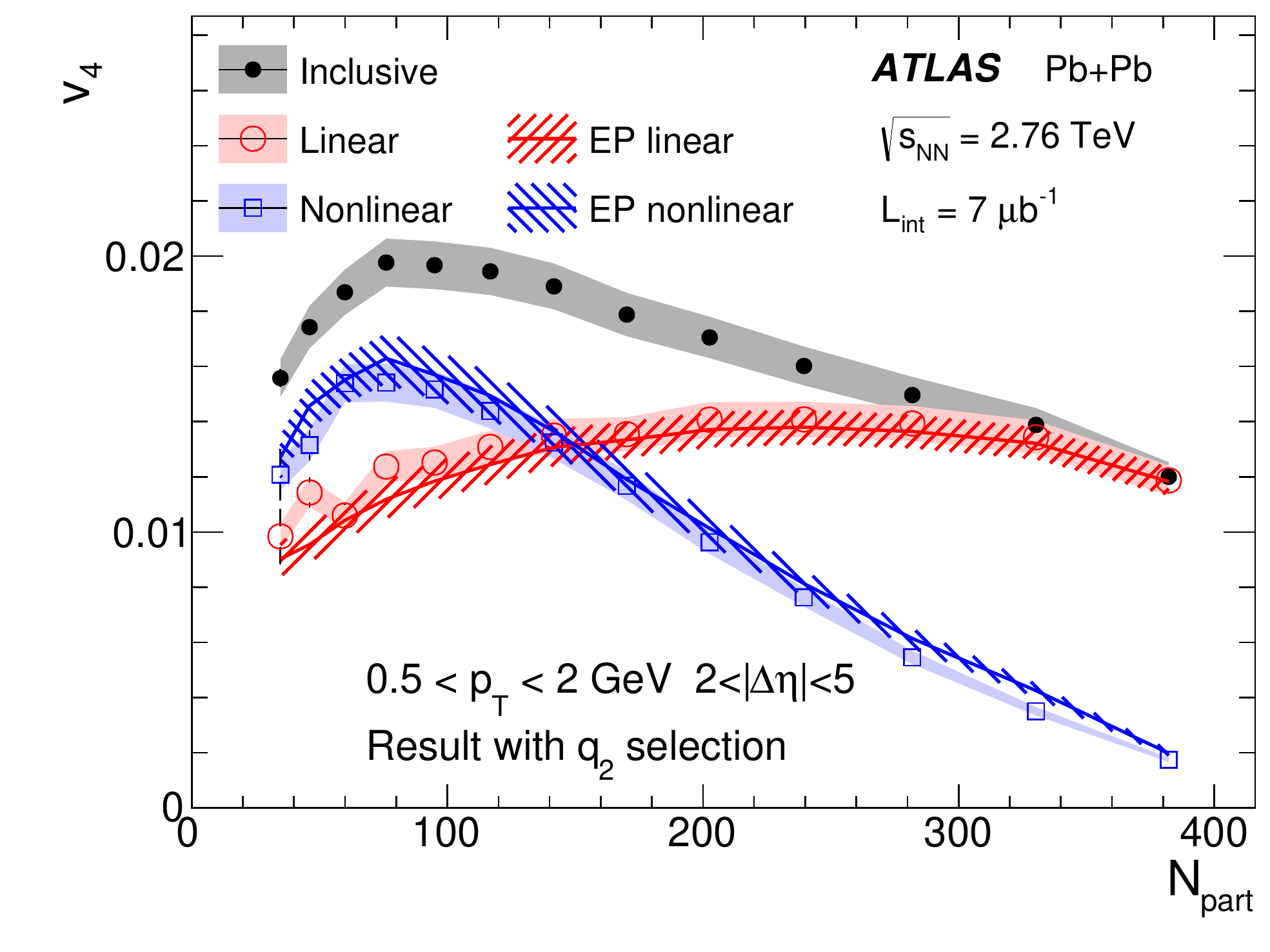}
\end{center}
\caption{\label{fig:3d} (Color online) The centrality ($\npart$) dependence of the $v_4$ in $0.5<\pT<2$~GeV and the associated linear and nonlinear components extracted from the fits in Fig.~\ref{fig:3c} and Eq.~(\ref{eq:2c}). They are compared with the linear and nonlinear component estimated from the previously published event-plane correlations~\cite{Aad:2014fla} via Eq.~(\ref{eq:2d}). The error bars represent the statistical uncertainties, while the shaded bands or hashed bands represent the systematic uncertainties.}
\end{figure}

The $c_0$ term from the fits can be used to decompose $v_4$, without $q_2$ selection, into linear and nonlinear terms for each centrality interval as:
\begin{eqnarray}
\label{eq:2c}
v_4^{\rm{L}}=c_0,\;\;\;\; v_4^{\rm{NL}}=\sqrt{v_4^2-c_0^2}\;.
\end{eqnarray}
The results as a function of centrality are shown in Fig.~\ref{fig:3d} (open circles and squares). The linear term associated with $\epsilon_4$ depends only weakly on centrality, and becomes the dominant part of $v_4$ for $\npart>150$, or 0--30\% centrality range. The nonlinear term increases as the collisions become more peripheral, and becomes the dominant part of $v_4$ for $\npart<$ 120.

Since the contributions of higher-order nonlinear terms are small, as suggested by the fits discussed above, the linear and nonlinear contributions can also be estimated directly from the previously published event-plane correlator $\left\langle\cos4(\Phi_2-\Phi_4)\right\rangle$ from ATLAS~\cite{Aad:2014fla}:
\begin{eqnarray}
\label{eq:2d}
v_4^{\rm{NL,EP}}=v_4\left\langle\cos4(\Phi_2-\Phi_4)\right\rangle,\;\;\;\; v_4^{\rm{L,EP}}= \sqrt{v_4^2-(v_4^{\rm{NL,EP}})^2}\;.
\end{eqnarray}
Results for this decomposition are shown in Fig.~\ref{fig:3d} (the hashed bands labeled EP), and they agree with the result obtained from the present analysis.

\begin{figure}[!h]
\begin{center}
\includegraphics[width=1\columnwidth]{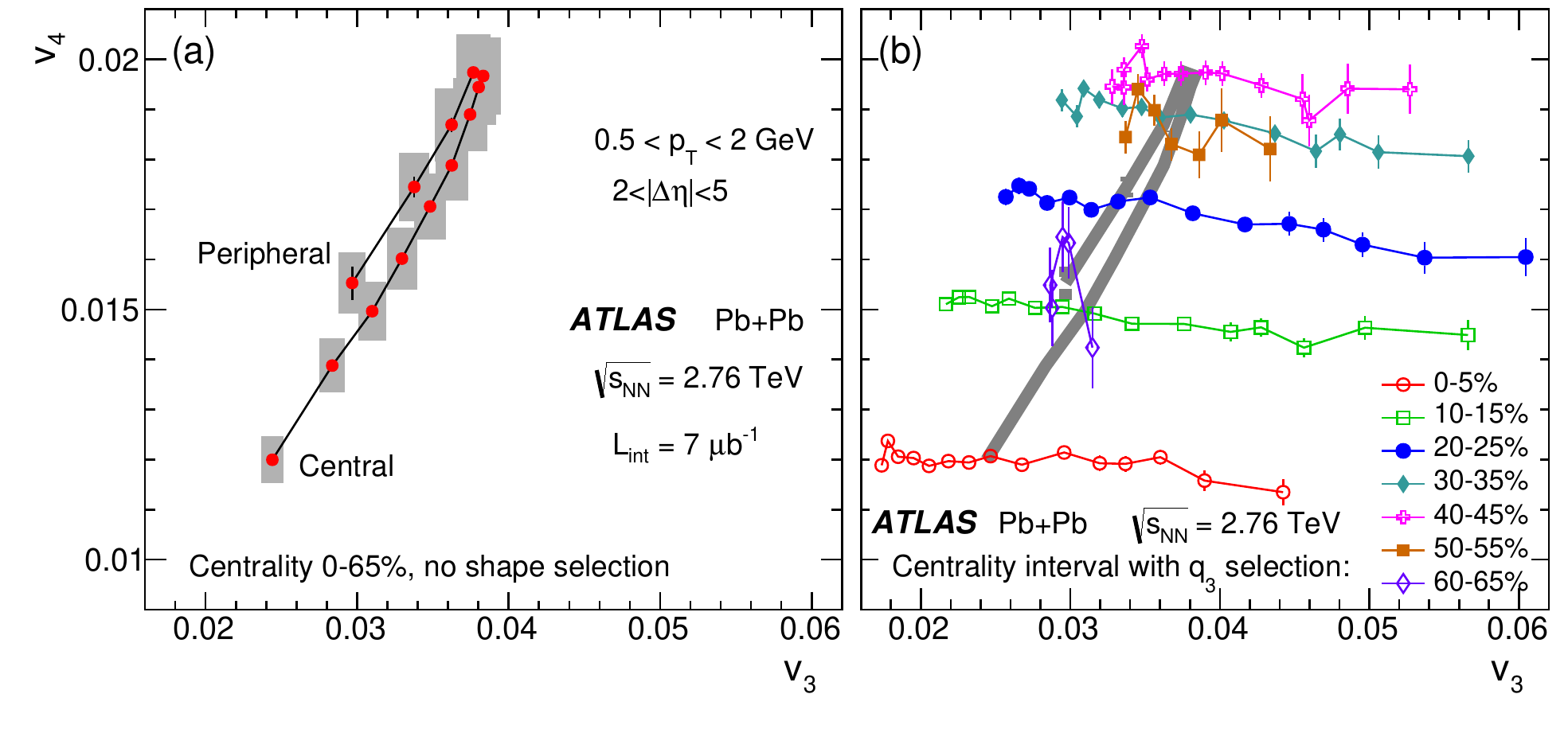}
\end{center}
\caption{\label{fig:3b} (Color online) The correlation of $v_3$ ($x$-axis) with $v_4$ ($y$-axis) both measured in $0.5<\pT<2$ GeV. The left panel shows the $v_3$ and $v_4$ values in thirteen 5\% centrality intervals over the centrality range 0--65\% without event-shape selection. The data points are connected to show the boomerang trend from central to peripheral collisions as indicated. The right panel shows the $v_3$ and $v_4$ values for fourteen $q_3$ selections (the two highest $q_3$ intervals in Fig.~\ref{fig:0a} are combined) in several centrality ranges (markers) with larger $v_3$ value corresponding to larger $q_3$ value; they are overlaid with the centrality dependence from the left panel. The error bars and shaded boxes represent the statistical and systematic uncertainties, respectively.}
\end{figure}

Figure~\ref{fig:3b}(a) shows the correlation between $v_3$ and $v_4$ in $0.5<\pT<2$~GeV prior to the event-shape selection. The data fall nearly on a single curve, reflecting the similar centrality dependence trends for $v_3$ and $v_4$~\cite{Aad:2012bu}. Figure~\ref{fig:3b}(b) shows the $v_3$--$v_4$ correlation for different $q_3$ event classes (colored symbols) overlaid with the centrality dependence taken from Fig.~\ref{fig:3b}(a) (thick solid line). A slight anticorrelation between $v_3$ and $v_4$ is observed, which is consistent with the fact that $v_4$ has a large nonlinear contribution from $v_2$ (Fig.~\ref{fig:3d}), which in turn is anticorrelated with $v_3$ (Fig.~\ref{fig:2}). 

\begin{figure}[!t]
\begin{center}
\includegraphics[width=1\columnwidth]{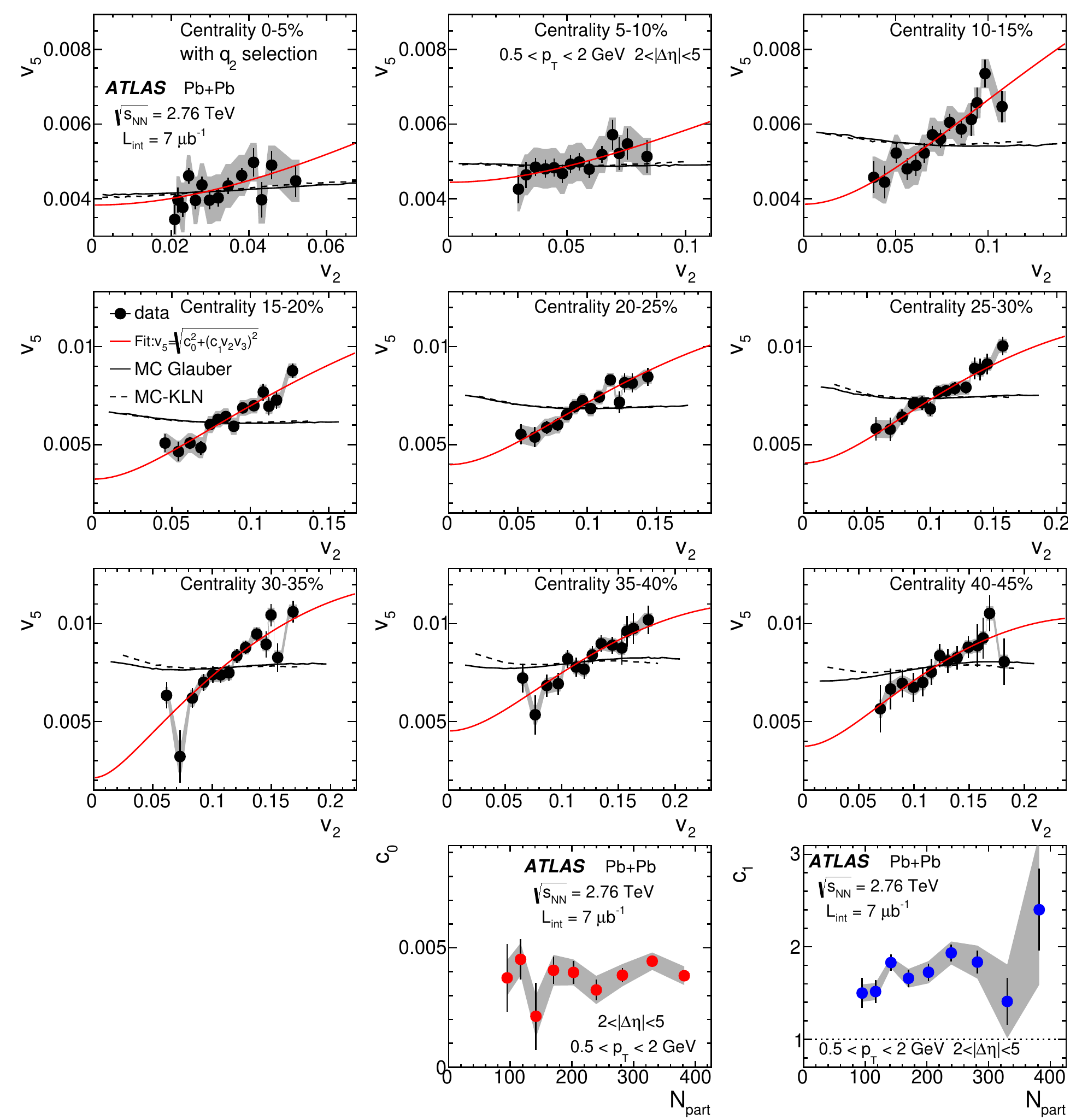}
\end{center}
\caption{\label{fig:4c} (Color online) The correlation of $v_2$ ($x$-axis) with $v_5$ ($y$-axis) in $0.5<\pT<2$~GeV for fourteen $q_2$ selections (the two highest $q_2$ intervals in Fig.~\ref{fig:0a} are combined) in nine 5\% centrality intervals. The data are compared with the rescaled $\epsilon_2$--$\epsilon_5$ correlation from MC Glauber and MC-KLN models in the same centrality interval. The data are also parameterized with Eq.~(\ref{eq:2e}), taking into account both statistical and systematic uncertainties. The $\npart$ dependence of the fit parameters is shown in the last two panels. The error bars and shaded bands represent the statistical and systematic uncertainties, respectively.}
\end{figure}
\begin{figure}[!h]
\begin{center}
\includegraphics[width=1\columnwidth]{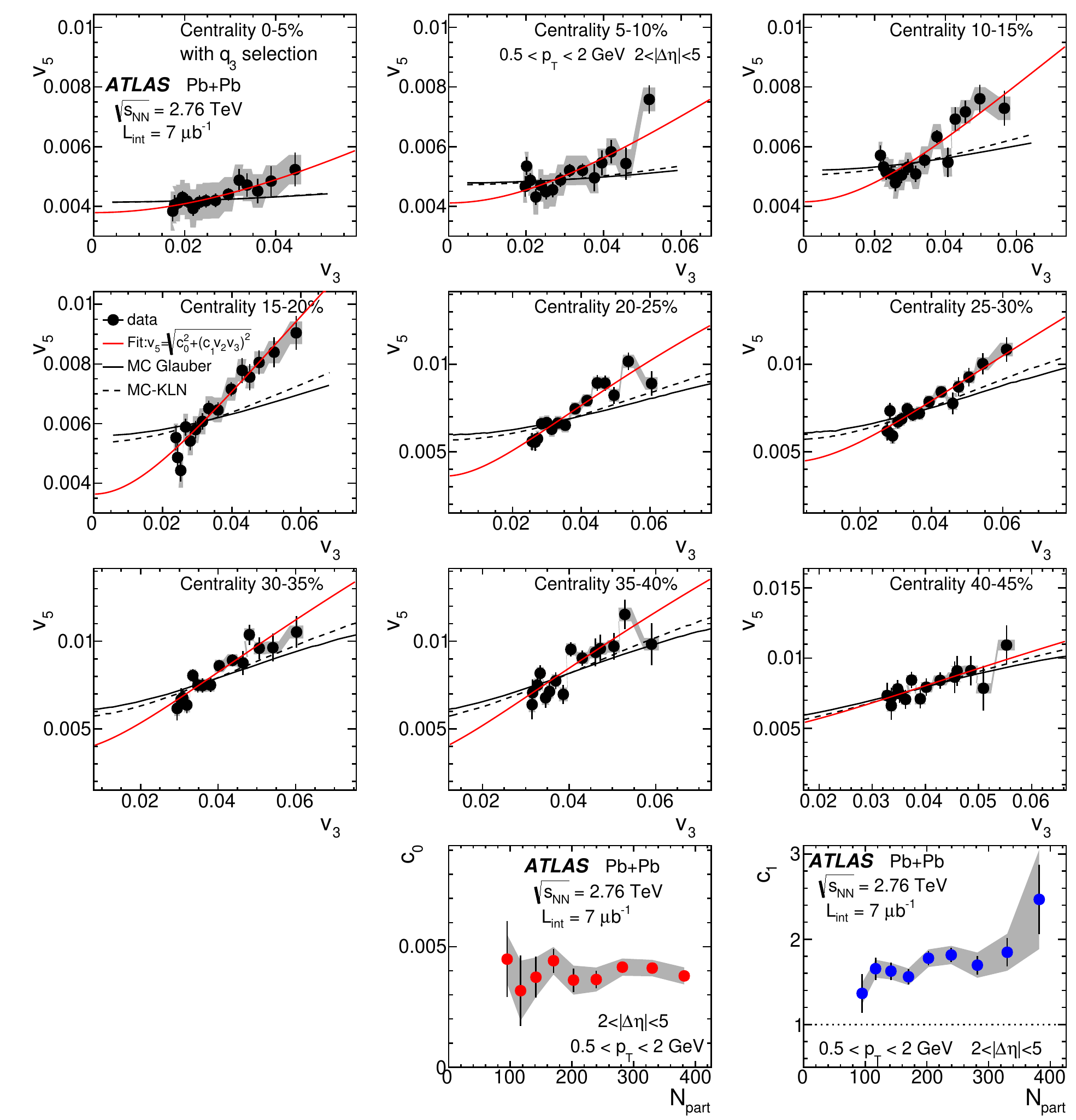}
\end{center}
\caption{\label{fig:4cc} (Color online) The correlation of $v_3$ ($x$-axis) with $v_5$ ($y$-axis) in $0.5<\pT<2$~GeV for fourteen $q_3$ selections (the two highest $q_3$ intervals in Fig.~\ref{fig:0a} are combined) in nine 5\% centrality intervals. The data are compared with the rescaled $\epsilon_3$--$\epsilon_5$ correlation from MC Glauber and MC-KLN models in the same centrality interval. The data are also parameterized with Eq.~(\ref{eq:2e}), taking into account both statistical and systematic uncertainties. The $\npart$ dependence of the fit parameters is shown in the last two panels. The error bars and shaded bands represent the statistical and systematic uncertainties, respectively.}
\end{figure}

\subsection{$v_2$--$v_5$ and $v_3$--$v_5$ correlations}
\label{sec:res4} 
The analysis of $v_2$--$v_5$ and $v_3-v_5$ correlations proceeds in the same manner as for the $v_2$--$v_4$ and $v_3$--$v_4$ correlations. A separation of the linear and nonlinear components of $v_5$ is made. 

Figure~\ref{fig:4c} shows the $v_2$--$v_5$ correlation in $0.5<\pT<2$~GeV with $q_2$ selection, separately for each centrality interval. The data are compared with the $\epsilon_2$--$\epsilon_5$ correlations rescaled according to Eq.~(\ref{eq:2aa}). The rescaled $\epsilon_2$--$\epsilon_5$ correlations fail to describe the data in all centrality intervals, suggesting that the nonlinear contribution in Eq.~(\ref{eq:0f}) is important. To separate the linear and nonlinear component in the $v_2$--$v_5$ correlation, the data are fitted with the following function:
\begin{eqnarray}
\label{eq:2e}
v_5&=& \sqrt{c_{0}^2 + (c_{1}v_2v_3)^{2}}\;,
\end{eqnarray} 
where the higher-order nonlinear terms in Eq.~(\ref{eq:0f}) and a possible cross-term associated with $\langle\cos(2\Psi_2+3\Psi_3-5\Psi_5)\rangle$ are dropped. For each centrality interval, Eq.~(\ref{eq:2a}) is used to fix the $v_3$ value for each $v_2$ value. The fits are shown in Fig.~\ref{fig:4c} and describe the data well for all centrality intervals. The centrality ($\npart$) dependence of the fit parameters is shown in the last two panels of Fig.~\ref{fig:4c}. The $c_0$ represents an estimate of the linear component of $v_5$, and the nonlinear term is driven by $c_1$, which has a value of $\sim1.5$--2.

Figure~\ref{fig:4cc} shows the $v_3$--$v_5$ correlations with $q_3$ selection in various centrality intervals. If Eq.~(\ref{eq:2e}) is a valid decomposition of $v_5$, then it should also describe these correlations. Figure~\ref{fig:4cc} shows that this indeed is the case. The parameters extracted from a fit to Eq.~(\ref{eq:2e}) as shown in the last two panels of Fig.~\ref{fig:4cc} are consistent with those obtained from $v_2$--$v_5$ correlations. 

\begin{figure}[!t]
\begin{center}
\includegraphics[width=0.5\columnwidth]{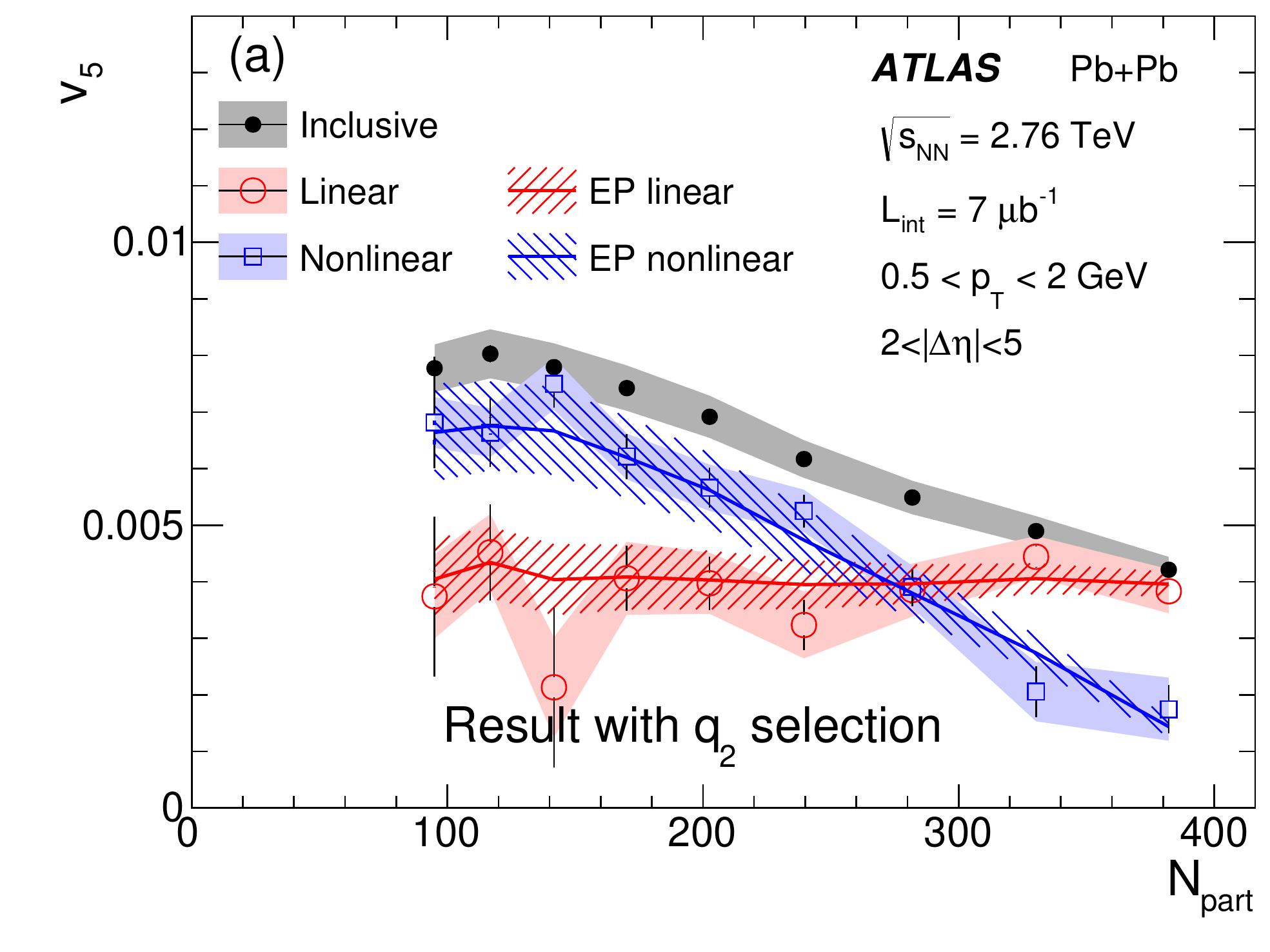}\includegraphics[width=0.5\columnwidth]{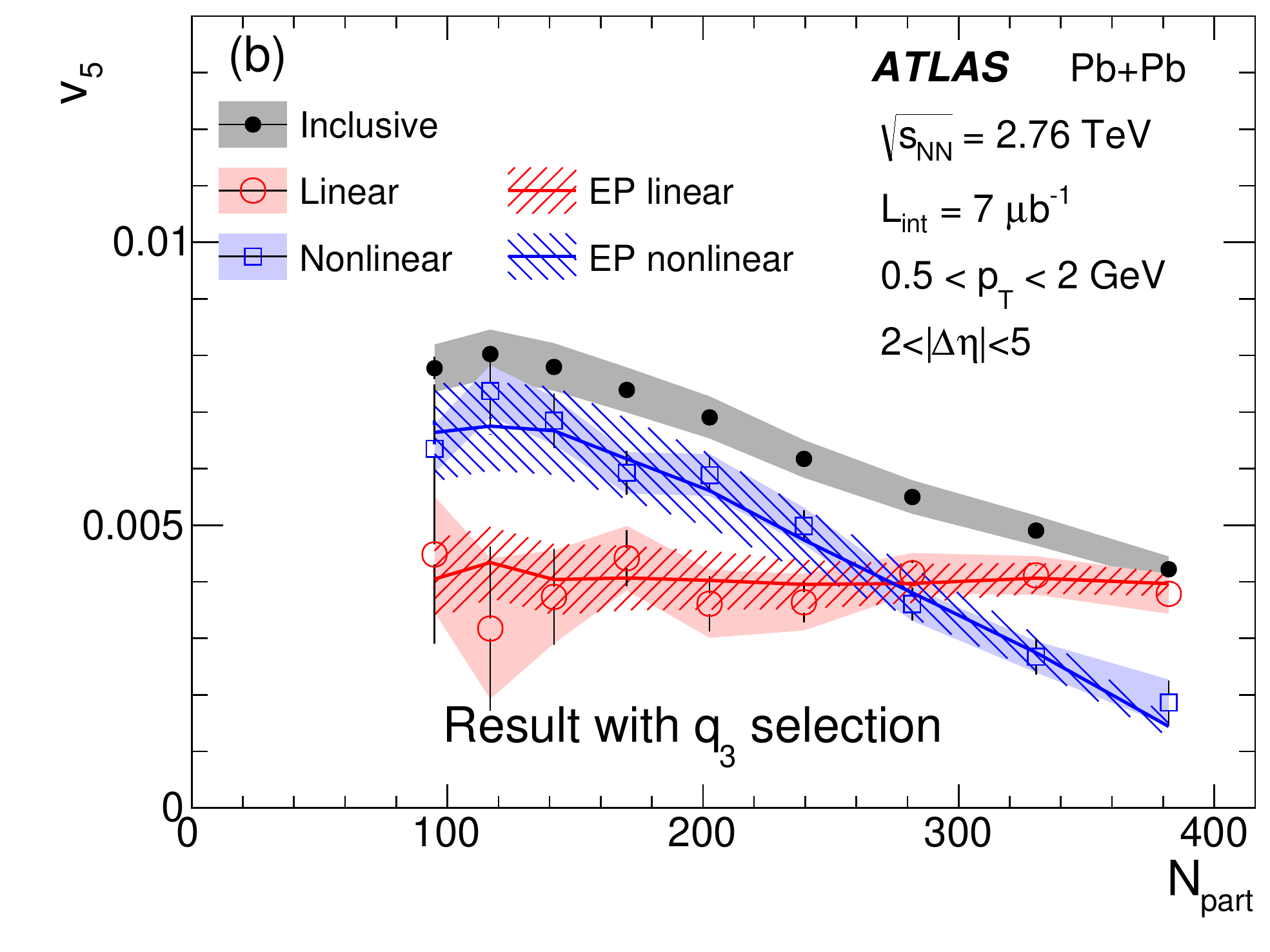}
\end{center}
\caption{\label{fig:4d} (Color online) The centrality ($\npart$) dependence of the $v_5$ in $0.5<\pT<2$~GeV and the associated linear and nonlinear components extracted from the fits in Fig.~\ref{fig:4c}, Fig.~\ref{fig:4cc} and Eq.~(\ref{eq:2f}). They are compared with the linear and nonlinear component estimated from the previous published event-plane correlation~\cite{Aad:2014fla} via Eq.~(\ref{eq:2g}). The error bars represent the statistical uncertainties, while the shaded bands or hashed bands represent the systematic uncertainties.}
\end{figure}

From the fit results in Figs.~\ref{fig:4c} and \ref{fig:4cc}, the inclusive $v_5$ values prior to event-shape selection are decomposed into linear and nonlinear terms for each centrality interval as:
\begin{eqnarray}
\label{eq:2f}
v_5^{\rm{L}}=c_0,\;\;\;\; v_5^{\rm{NL}}=\sqrt{v_5^2-c_0^2}\;.
\end{eqnarray}
The results as a function of centrality are shown in the two panels of Fig.~\ref{fig:4d}, corresponding to Fig.~\ref{fig:4c} and \ref{fig:4cc}, respectively. Results for the two decompositions show consistent centrality dependence: the linear term associated with $\epsilon_5$ dominates the $v_5$ signal only in the most central collisions ($\npart>300$ or 0--10\% centrality). The nonlinear term increases as the collisions become more peripheral, and becomes the dominant part of $v_5$ for $\npart\lesssim280$. 

Similar to the case of $v_2$--$v_4$ correlation, the linear and nonlinear contribution to $v_5$ can also be estimated directly from the previously published event-plane correlator $\left\langle\cos(2\Phi_2+3\Phi_3-5\Phi_5)\right\rangle$ from ATLAS~\cite{Aad:2014fla}:
\begin{eqnarray}
\label{eq:2g}
v_5^{\rm{NL,EP}}=v_5\left\langle\cos(2\Phi_2+3\Phi_3-5\Phi_5)\right\rangle, \;\;\;\;v_5^{\rm{L,EP}}= \sqrt{v_5^2-(v_5^{\rm{NL,EP}})^2}\;.
\end{eqnarray}
Results for this decomposition are shown as solid curves in Fig.~\ref{fig:4d}, and they agree well with the result obtained in the present analysis.

\subsection{Eccentricity-scaled $v_n$}
\label{sec:res5} 
\begin{figure}[h]
\begin{center}
\includegraphics[width=1\columnwidth]{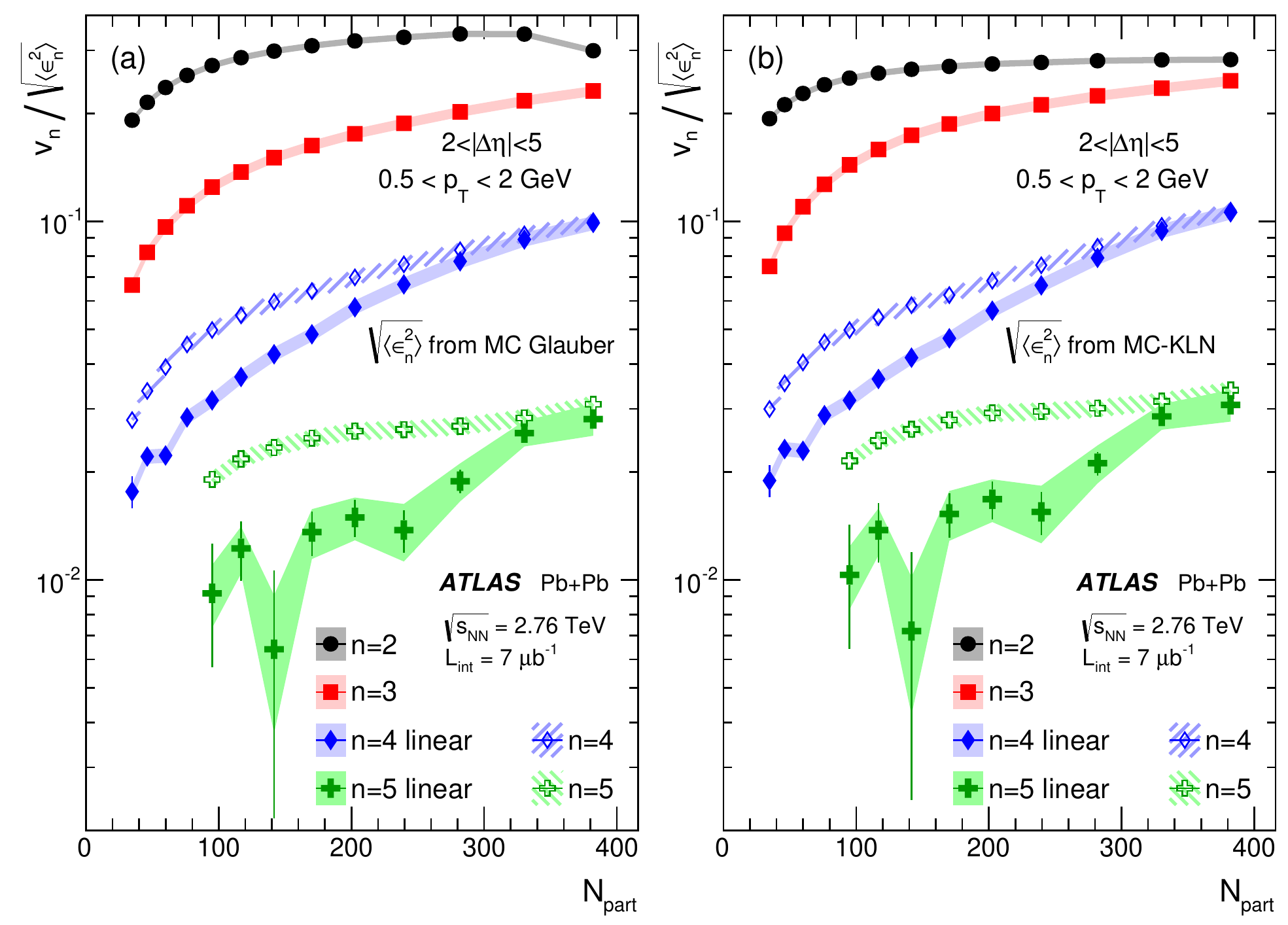}
\end{center}
\caption{\label{fig:6} (Color online) The eccentricity-scaled $v_n$ or the estimated linear component $v_n^{\rm L}$ obtained from two-component fits, $v_2/\sqrt{\langle\epsilon_2^2\rangle}$ (circles),  $v_3/\sqrt{\langle\epsilon_3^2\rangle}$ (boxes), $v_4^{\rm L}/\sqrt{\langle\epsilon_4^2\rangle}$ (solid diamonds),  $v_5^{\rm L}/\sqrt{\langle\epsilon_5^2\rangle}$ (solid crosses),  $v_4/\sqrt{\langle\epsilon_4^2\rangle}$ (open diamonds), and $v_5/\sqrt{\langle\epsilon_5^2\rangle}$ (open crosses). The eccentricities are calculated from the MC Glauber model (left panel) and the MC-KLN model (right panel). The error bars represent the statistical uncertainties, while the shaded bands or hashed bands represent the systematic uncertainties.}
\end{figure}

One quantity often used to characterize the collective response of the medium to the initial geometry is the response coefficient $k_n$ defined in Eq.~(\ref{eq:0dd}). Since the $v_n$ obtained from the two-particle correlation method effectively measure the RMS values of the event-by-event $v_n$~\cite{Adcox:2002ms}, a more appropriate quantity to characterize the collective response is the ratio of $v_n$ to the RMS eccentricity~\cite{Voloshin:2008dg,Qiu:2011iv}: $v_n/\sqrt{\langle\epsilon_n^2\rangle}$. This quantity can be directly calculated for $v_2$ and $v_3$ since they are mostly driven by $\epsilon_2$ and $\epsilon_3$. But for higher-order flow harmonics, it is more appropriate to use the extracted linear component $v_n^{\rm L}$ to make the ratios as it is more directly related to the $\epsilon_n$. The $v_n^{\rm L}$ is taken as the $c_0$ term obtained from the two-component fits in Fig.~\ref{fig:3c} for $n=4$ and Fig.~\ref{fig:4c} for $n=5$. Figure~\ref{fig:6} shows the centrality dependence of $v_n/\sqrt{\langle\epsilon_n^2\rangle}$ for $n=2$ and 3 and $v_n^{\rm L}/\sqrt{\langle\epsilon_n^2\rangle}$ for $n=4$ and 5 (denoted by ``linear'' in figure legend), with $\epsilon_n$ calculated in the MC Glauber model (left panel) and MC-KLN model (right panel). The higher-order flow harmonics show increasingly strong centrality dependence, which is consistent with the stronger viscous-damping effects as expected from hydrodynamic model calculations~\cite{Teaney:2010vd,Alver:2010dn,Schenke:2011bn}. For comparison, the ratios are also shown for the full $v_4$ and $v_5$ values without the linear and nonlinear decomposition, i.e., $v_4/\sqrt{\langle\epsilon_4^2\rangle}$ (open diamonds) and $v_5/\sqrt{\langle\epsilon_5^2\rangle}$ (open crosses); they show much weaker centrality dependence due to the dominance of nonlinear contributions to more peripheral collisions.
%\clearpage
\section{Conclusion}
Correlations between $v_m$ coefficients for $m=2$ or 3 in different $\pT$ ranges, and the correlation between $v_m$ and other flow harmonics $v_n$ for $n=2$ to 5 in the same $\pT$ range, are presented using 7 $\mu$b$^{-1}$ of Pb+Pb collision data at $\sqrt{s_{_{N\!N}}}=2.76$~TeV collected in 2010 by the ATLAS experiment at the LHC. The $v_m$--$v_n$ correlations are measured for events within a given narrow centrality interval using an event-shape selection method. Beside the centrality selection, this method makes a further classification of events according to their raw elliptic flow signal $q_2$ or raw triangular flow signal $q_3$ in the forward rapidity range $3.3<|\eta|<4.8$. For each $q_m$ bin, the $v_m$ and $v_n$ coefficients are calculated at midrapidity $|\eta|<2.5$ using a two-particle correlation method, and hence the differential $v_m$--$v_n$ correlation within each centrality interval can be obtained. 

The correlation of $v_m$ between two different $\pT$ ranges shows a complex centrality dependence, but within a narrow centrality interval the correlation varies linearly with the event shape as determined by $q_2$ or $q_3$. This linearity indicates that the viscous effects are controlled by the system size not by its overall shape. An anticorrelation is observed between $v_2$ and $v_3$ within a given centrality interval, and agrees qualitatively with similar anticorrelation between corresponding eccentricities $\epsilon_2$ and $\epsilon_3$, indicating that these correlations are associated with initial geometry effects.

The $v_4$ is found to increase strongly with $v_2$, and $v_5$ is found to increase strongly with both $v_2$ and $v_3$ within a given centrality interval. The trends and the strengths of $v_2$--$v_4$, $v_2$--$v_5$, and $v_3$--$v_5$ correlations disagree with corresponding $\epsilon_m$--$\epsilon_n$ correlations predicted by MC Glauber and MC-KLN initial geometry models. Instead, these correlations are found to be consistent with a combination of a linear contribution to $v_4$ from $\epsilon_4$ and to $v_5$ from $\epsilon_5$, together with a nonlinear term that is a function of $v_2^2$ or of $v_2v_3$, as predicted by hydrodynamic models~\cite{Gardim:2011xv,Teaney:2012gu}. The functional form of these nonlinear contributions is eclipsed in the overall centrality dependence, but has been directly exposed in the event-shape selected measurements reported here. A simple two-component fit is used to separate these two contributions in $v_4$ and $v_5$. The extracted linear and nonlinear contributions are found to be consistent with those obtained from previously measured event-plane correlations.

In order to quantify the response of the medium to the initial geometry, the extracted linear components of $v_4$ and $v_5$,  $v_4^{\rm L}$ and $v_5^{\rm L}$, are scaled by the RMS eccentricity of corresponding order. The scaled quantities, $v_4^{\rm L}/\sqrt{\langle\epsilon_4^2\rangle}$ and $v_5^{\rm L}/\sqrt{\langle\epsilon_5^2\rangle}$, show stronger centrality dependence than the similarly scaled quantities for elliptic flow and triangular flow, consistent with the stronger viscous-damping effects expected for higher-order harmonics.

% Acknowledgements for papers with collision data
% Version 23-Mar-2015

% Standard acknowledgements start here
%----------------------------------------------
\section*{ACKNOWLEDGMENTS}
We thank CERN for the very successful operation of the LHC, as well as the
support staff from our institutions without whom ATLAS could not be
operated efficiently.

We acknowledge the support of ANPCyT, Argentina; YerPhI, Armenia; ARC,
Australia; BMWFW and FWF, Austria; ANAS, Azerbaijan; SSTC, Belarus; CNPq and FAPESP,
Brazil; NSERC, NRC and CFI, Canada; CERN; CONICYT, Chile; CAS, MOST and NSFC,
China; COLCIENCIAS, Colombia; MSMT CR, MPO CR and VSC CR, Czech Republic;
DNRF, DNSRC and Lundbeck Foundation, Denmark; EPLANET, ERC and NSRF, European Union;
IN2P3-CNRS, CEA-DSM/IRFU, France; GNSF, Georgia; BMBF, DFG, HGF, MPG and AvH
Foundation, Germany; GSRT and NSRF, Greece; RGC, Hong Kong SAR, China; ISF, MINERVA, GIF, I-CORE and Benoziyo Center, Israel; INFN, Italy; MEXT and JSPS, Japan; CNRST, Morocco; FOM and NWO, Netherlands; BRF and RCN, Norway; MNiSW and NCN, Poland; GRICES and FCT, Portugal; MNE/IFA, Romania; MES of Russia and NRC KI, Russian Federation; JINR; MSTD,
Serbia; MSSR, Slovakia; ARRS and MIZ\v{S}, Slovenia; DST/NRF, South Africa;
MINECO, Spain; SRC and Wallenberg Foundation, Sweden; SER, SNSF and Cantons of
Bern and Geneva, Switzerland; NSC, Taiwan; TAEK, Turkey; STFC, the Royal
Society and Leverhulme Trust, United Kingdom; DOE and NSF, United States of
America.

The crucial computing support from all WLCG partners is acknowledged
gratefully, in particular from CERN and the ATLAS Tier-1 facilities at
TRIUMF (Canada), NDGF (Denmark, Norway, Sweden), CC-IN2P3 (France),
KIT/GridKA (Germany), INFN-CNAF (Italy), NL-T1 (Netherlands), PIC (Spain),
ASGC (Taiwan), RAL (UK) and BNL (USA) and in the Tier-2 facilities
worldwide.
%----------------------------------------------

\bibliographystyle{atlasBibStyleWoTitle}
\providecommand{\href}[2]{#2}\begingroup\raggedright\endgroup

%\bibliography{hion-2014-03_v9}
\newpage % ATLAS Collaboration author list
% Data extracted on 27-Mar-2015 for paper reference HION-2014-03
%\documentclass[11pt]{article}
%\usepackage{a4wide}\begin{document}
\begin{flushleft}
{\Large The ATLAS Collaboration}

\bigskip

G.~Aad$^{\rm 85}$,
B.~Abbott$^{\rm 113}$,
J.~Abdallah$^{\rm 151}$,
O.~Abdinov$^{\rm 11}$,
R.~Aben$^{\rm 107}$,
M.~Abolins$^{\rm 90}$,
O.S.~AbouZeid$^{\rm 158}$,
H.~Abramowicz$^{\rm 153}$,
H.~Abreu$^{\rm 152}$,
R.~Abreu$^{\rm 30}$
% ...
\bigskip
\\
% ...
$^{11}$ Institute of Physics, Azerbaijan Academy of Sciences, Baku, Azerbaijan\\
% ...
$^{30}$ CERN, Geneva, Switzerland\\
% ...
$^{85}$ CPPM, Aix-Marseille Universit{\'e} and CNRS/IN2P3, Marseille, France\\
% ...
$^{90}$ Department of Physics and Astronomy, Michigan State University, East Lansing MI, United States of America\\
% ...
$^{107}$ Nikhef National Institute for Subatomic Physics and University of Amsterdam, Amsterdam, Netherlands\\
% ...
$^{113}$ Homer L. Dodge Department of Physics and Astronomy, University of Oklahoma, Norman OK, United States of America\\
% ...
$^{151}$ Institute of Physics, Academia Sinica, Taipei, Taiwan\\
$^{152}$ Department of Physics, Technion: Israel Institute of Technology, Haifa, Israel\\
$^{153}$ Raymond and Beverly Sackler School of Physics and Astronomy, Tel Aviv University, Tel Aviv, Israel\\
% ...
$^{158}$ Department of Physics, University of Toronto, Toronto ON, Canada\\
% ...
\end{flushleft}

%\end{document}
% Created with ./xml2latex.py
 
\end{document}